\documentclass[12pt]{article}
\usepackage[letterpaper,margin=1in]{geometry}
\usepackage{setspace}
\usepackage{hyperref}
\usepackage{natbib}
\usepackage{amsmath,mathrsfs,amsthm,graphicx,bm,amssymb}
\usepackage{algorithm}
\usepackage{color}
\usepackage{ulem}                           
\usepackage{enumitem}
\usepackage{multirow}
\usepackage{pdflscape}
\usepackage{bbm}

\newtheorem{theorem}{Theorem}[subsection]
\newtheorem{lemma}{Lemma}[subsection]
\newtheorem{definition}{Definition}[section]
\newtheorem{remark}{Remark}[subsection]
\DeclareMathOperator*{\argmin}{argmin}

\newcommand{\blind}{1}
\begin{document}

\def\spacingset#1{\renewcommand{\baselinestretch}%
{#1}\small\normalsize} \spacingset{1}

\setlength{\belowdisplayskip}{0pt} \setlength{\belowdisplayshortskip}{0pt}
\setlength{\abovedisplayskip}{0pt} \setlength{\abovedisplayshortskip}{0pt}

\if1\blind
{
  \title{\bf High-dimensional Measurement Error Models for Lipschitz Loss}
\author{
Xin Ma\footnote{Department of Statistics, Florida State University} and Suprateek Kundu\footnote{Department of Biostatistics, The University of Texas at MD Anderson Cancer Center} \footnote{Corresponding author: Email: SKundu2@mdanderson.org; Address: 1400 Pressler Street, Unit 1411, Houston, TX 77030}
}
  \date{}
  \maketitle
} \fi

\if0\blind
{
  \bigskip
  \bigskip
  \bigskip
  \begin{center}
    {\LARGE\bf High-dimensional Measurement Error Models for Lipschitz Loss}
\end{center}
  \medskip
} \fi

\begin{abstract}
Recently emerging large-scale biomedical data pose exciting opportunities for scientific discoveries. However, the ultrahigh dimensionality and non-negligible measurement errors in the data may create difficulties in estimation.  There are limited methods for high-dimensional covariates with measurement error, that usually require knowledge of the noise distribution and focus on linear or generalized linear models. In this work, we develop high-dimensional measurement error models for a class of Lipschitz loss functions that encompasses logistic regression, hinge loss and quantile regression, among others.  Our estimator is designed to minimize the $L_1$ norm among all estimators belonging to suitable feasible sets, without requiring any knowledge of the noise distribution. Subsequently, we generalize these estimators to a Lasso analog version that is computationally scalable to higher dimensions. We derive theoretical guarantees in terms of finite sample statistical error bounds and sign consistency, even when the dimensionality increases exponentially with the sample size. Extensive simulation studies demonstrate superior performance compared to existing methods in classification and quantile regression problems. An application to a gender classification task based on brain functional connectivity in the Human Connectome Project data illustrates improved accuracy under our approach, and the ability to reliably identify significant brain connections that drive gender differences.
\end{abstract}
\noindent%
{\it Keywords:} Classification; Lipschitz loss; measurement error models; neuroimaging analysis.
\vfill

\spacingset{1.5} 

\section{Introduction}
\label{sec:intro}
High-dimensional data has emerged in various research fields such as human genetics, neuroimaging, and microbiome studies. When the number of features in the data become larger than the sample size or even increases exponentially with the sample size, the traditional regression models would fail to provide an estimation for the regression coefficients, and the theoretical large sample results would not apply. In order to accommodate the ultra high number of features in the regression framework, a series of penalized methods have been proposed. These methods assume that there are only a small set of features contributing to the outcome variable, thus the regression coefficients only include very few nonzero elements. Well-known examples of the sparse learning methods include Lasso with convex $L_1$ penalty \citep{tibshirani1996regression} and the closely related Dantzig selector \citep{bickel2009simultaneous}, and the non-convex type of methods such as the smoothly clipped absolute deviation (SCAD) \citep{fan2001variable} and the minimax concave penalty (MCP) \citep{zhang2010nearly}, among others.

For these penalized methods, the sparse assumption on the true regression coefficients have enabled establishment of desirable theoretical results, both for linear regression models as well as generalized linear regression settings \citep{james2009generalized}. More recently, \cite{negahban2012unified} proposed a unified framework for M-estimators based on decomposable regularizers for a wide class of convex differentiable loss functions. In the context of generalized linear models, \cite{fan2011nonconcave} investigated the performance of specific non-convex penalties including SCAD and MCP in ultrahigh dimensions and showed them to possess the oracle property under mild assumptions. For the specific case of binary classification in the context of penalized support vector machines (SVM), \cite{peng2016error} derived the finite sample statistical error bounds under $L_1$ penalties and further showed the oracle property of non-convex penalized SVM under certain conditions. A wider class of Lipschitz loss functions involving support vector machines and quantile regression was studied by \cite{dedieu2019sparse} under varying penalties with a focus on deriving the finite sample statistical error bounds under high-dimensional settings.

The above approaches, and the traditional literature on penalized approaches for high-dimensional regression, has essentially ignored the presence of measurement error in high-dimensional covariates. However, this can be an unrealistic assumption, especially in medical imaging studies, where measurement errors in the images can result due to various factors such as technical limitations and experimental design, measurement errors and so on \citep{raser2005noise,liu2016noise}. Ignoring measurement error in estimation has been shown to result in biased estimates and
attenuation to the null \citep{carroll1994measurement}. A more recent line of work has extended penalized sparse learning to the case of high dimensional covariates with measurement errors in linear regression settings.  See for example, recent work by \cite{loh2012high,datta2017cocolasso} and more recent work involving grouped penalties in the presence of noisy functional features \citep{ma2022multi}. These approaches typically used corrected versions of the objective functions in order to tackle the measurement error in covariates. Such approaches often require additional validation samples in order to compute moments of the measurement error distribution, which may not always be feasible in practice. Further, storing the high-dimensional noise covariance may impose excessive  memory requirements that can result in computational bottlenecks. Some recent approaches bypass the challenges with computing and storing the high-dimensional noise covariance, by not requiring knowledge of the noise distributions. Examples include the seminal work by  \cite{rosenbaum2010sparse,rosenbaum2013improved} involving the matrix uncertainty selector (MUS) that is motivated by the Dantzig selector, and relies on feasible sets that is guaranteed to contain the true parameters with high confidence. Other examples include sparse total least squares \citep{zhu2011sparsity}, and orthogonal matching pursuit \citep{chen2013noisy}.

There has been a parallel development of generalized linear models involving measurement errors in covariates, which has unfortunately been restricted to fixed or low dimensional covariate features.  Readers can refer to \cite{stefanski2000measurement} for a review on measurement error models. In a recent work,  \cite{sorensen2018covariate} proposed a heuristic approach known as the generalized MUS (GMUS) estimator for generalized linear models (GLM); however, no theoretical aspects in terms of error bounds were investigated. Beyond the GLM and linear regression, there are very limited approaches for measurement error models, to our knowledge. Some examples include quantile regression methods involving covariates with noise such as \cite{wei2009quantile} who proposed a bias-correction approach based on joint estimating equations, and \cite{wang2012corrected} who proposed a method based on corrected score functions. The above approaches for GLM and quantile regression  typically have desirable asymptotic properties, but theoretical guarantees in terms of finite sample error bounds for these cases are lacking and they do not cater to high-dimensional settings of interest where the number of covariates may increase exponentially with the sample size. In this context, we note that deriving provably flexible estimators for high-dimensional covariates ($n<<p$) in classification as well as quantile-regression scenarios is known to be a challenging problem, even in the case without measurement errors \citep{peng2016error,dedieu2019sparse}.

In this article, we address the task of developing provably flexible estimators for high-dimensional covariates in the presence of measurement errors, for a general class of Lipschitz continuous loss functions that go beyond the routinely studied simple linear regression and GLM settings. In addition to logistic regression that lies within the GLM family, the class of loss functions include support vector machine (SVM) classification and quantile regression loss problems as special cases, among others. We note that a similar class of loss functions was investigated in \cite{dedieu2019sparse}, who proposed a $L_1$ constrained slope estimation but without considering measurement errors in covariates. Our work is distinct compared to their approach, both in terms of methodology as well as theoretical derivations and implementation. Our estimators are designed to minimize the $L_1$ norm among a wide class of estimators belonging to suitable feasible sets that contain the true parameter with high confidence. The proposed approach is motivated by developments in 
\cite{rosenbaum2010sparse} who proposed matrix uncertainty selectors (MUS) for linear regression problems in the presence of measurement error. However, our contributions constitute non-trivial generalizations of the MUS to a much wider class of loss functions that cover commonly encountered classification and quantile regression problems as special cases. Another distinction compared to \cite{rosenbaum2010sparse} is the fact that we propose a lasso analog of the proposed method that is computationally scalable to much higher dimensions and establish finite sample error bounds for this analog estimator. In our treatment, we start with the case without measurement error, and subsequently generalize our framework to the case with additive measurement errors. We will denote the proposed estimators for the general class of smooth Lipschitz loss functions as matrix uncertainty estimators for Lipschitz loss (MULL) throughout the article. To our knowledge, the proposed development addresses a critical gap in high-dimensional measurement error modeling literature.

We derive the finite sample statistical error bounds and sign consistency results for the proposed MULL estimators and the corresponding Lasso analog version, even when the number of covariates increase exponentially with the sample size, which guarantee sound operating characteristics of the method in high-dimensional cases. We implement the proposed approaches via efficient algorithms that scale to high-dimensional settings. Similar to \cite{rosenbaum2010sparse}, the proposed approach does not require  additional validation samples to compute the moments of the measurement error distribution that helps alleviate practical difficulties. We evaluate the operating characteristics via extensive simulation experiments and illustrate the superior performance of our proposed estimator over other competing methods in a variety of settings. The proposed approach is applied to a gender classification task based on functional connectome in the Human Connectome Project (HCP), where the number of edges in the brain network increases quadratically with the number of brain regions. In these types of studies, the brain network is estimated from the observed functional magnetic resonance imaging (fMRI) data that is noisy and hence subject to measurement error. Unfortunately, recent classification and regression approaches involving brain connectome have essentially ignored the presence of noise in the computed networks  \citep{guha2021bayesian,ma2022semi,relion2019network}. The proposed approach illustrates considerably higher classification accuracy compared to existing approaches. In addition, it demonstrates high test-retest reliability of the selected network features that is orders or magnitude higher than competing methods without or with noise correction.

We make several significantly novel contributions in this article: (i) develop an unifying penalized regression approach for high-dimensional covariates with measurement error under a broad class of Lipschitz continuous loss functions; (ii) propose a lasso analog version of the proposed approach that is more computationally scalable to high-dimensional covariates; (iii) derive theoretical guarantees for both the proposed variants in terms of finite sample error bounds as well as sign consistency; and (iv) develop an efficient computational algorithm for implementation of the above approaches. To our knowledge, this work is one of the first to develop a  provably flexible approach involving noisy  high-dimensional covariates and for a class of loss functions that go beyond the GLM family and linear models.

We give the outline of the rest of the article here. In Section \ref{sec:method}, we define our model framework and present the theoretical results. In Section \ref{sec:compute}, we develop the computational algorithm and provide advice on parameter tuning and initialization. Section \ref{sec:sims} reports results from extensive simulation studies.  Section \ref{sec:application} applies the proposed approach to HCP neuroimaging data. Section \ref{sec:discuss} provides discussion and future directions.

\vspace{-.25in}
\section{Proposed Methods with Lipschitz Losses}
\label{sec:method}
We consider a dataset with independent samples $\mathcal{D}_n := \{(\bm{x}_i,y_i)\in\mathbb{R}^p\times\mathcal{Y}, i=1,\cdots,n\}$, where the relationship between the outcome $y\in \mathcal{Y}$ and covariates $\bm{x}$ are defined by a pre-specified class of loss functions denoted by $f(\cdot,y)$. As elaborated below, the class of general loss functions considered in this article can encompass both binary classification and quantile regression problems, which corresponds to $y$ being discrete or continuous respectively. Given a pre-specified loss function $f(\cdot,y)$, we define the true coefficient $\bm{\beta}^*$ as the one that minimizes the theoretical loss function. In other words,  $\bm{\beta}^* = \argmin_{\bm{\beta}}\{E(\mathcal{L}(\bm{\beta}))\} = \argmin_{\bm{\beta}}\{E(f(\langle \bm{x},\bm{\beta}\rangle; y))\}$, where $\mathcal{L}(\bm{\beta}) = (1/n)\sum_{i=1}^n f(\langle \bm{x}_i,\bm{\beta}\rangle; y_i)$ denotes the empirical loss function, and  noting that the expectation is over the joint distribution of $\bm{x}$ and $y$. It will be assumed throughout that $||\bm{\beta}^*||_1\le R^*$, where $||\cdot||_1$ refers to $L_1$ norm. Our goal is to study the theoretical and empirical properties under such loss functions in the presence of  high-dimensional covariates with measurement errors. The high-dimensional settings considered in our work corresponds to ultra-high dimensions where the number of covariates $p$ is allowed to grow  exponentially with $n$ such that $\log(p)/n$ goes towards zero with increasing $n$ and $p$. Such settings are typically encountered in our motivating neuroimaging applications, where the number of voxels in a brain image can be orders of magnitudes higher than the sample size, or brain network-based analysis where the number of candidate edges in the network vastly exceeds the sample size. In such high dimensional settings, we typically assume that the true coefficient $\bm{\beta}^*$ to be sparse (involves $k<<p$ non-zero coefficients), which is routinely done in literature. Under this set-up, our goal is to find a sparse estimator $\hat{\bm{\beta}}$ that is close to $\bm{\beta}^*$ for a general class of loss functions that satisfy certain reasonable conditions.

The class of loss functions considered in our paper is assumed to be Lipschitz continuous that admits first and second-order derivatives as elaborated in the following Definition \ref{def:def1}. As illustrated in the sequel, this class of loss functions involve several commonly used losses. 

\begin{definition}
\label{def:def1}
A non-negative, convex loss function $f(\cdot,y)$ is L-Lipschitz continuous if 
$|f(t_1,y)-f(t_2,y)|\le L|t_1-t_2|, \forall t_1,t_2,$
and there exists first-order derivative function $\partial f(\cdot,y)$ such that 
$f(t_2,y)-f(t_1,y)\ge \partial f(t_1,y)(t_2-t_1), \forall t_1,t_2.$
In addition, the loss function is assumed to be twice-differentiable and admits a second-order derivative function $\partial^2 f(\cdot,y)$.
\end{definition}

We note that the derivatives are with respect to the first argument of the loss function. The above class includes commonly used loss functions for classification such as logistic regression that takes the form $f(t;y) = -yt + \log\{1+e^t\}$, and the smooth hinge loss that is an adaptive version of the original hinge loss for support vector machine (SVM) and takes the form $f(t;y) = \frac{1}{2}(1-yt) + \frac{1}{2}\sqrt{(1-yt)^2+\sigma^2}$, where $\mathcal{Y} = \{-1,1\}$, $\sigma>0$. It also includes a smoothed version of the quantile loss, named conquer loss, which is twice-differentiable and globally convex  \citep{he2021smoothed}. Here, $\mathcal{Y} = \mathbb{R}$ and the loss function takes the form $f(t;y) = l_h(y - t)$ with $l_h(u) = (\rho_\tau * K_h)(u) = \int_{-\infty}^\infty \rho_\tau(v)K_h(v-u)\mathrm{d}v$ where $*$ denotes the convolution operator, $\rho_\tau(u) = u\{\tau - \mathbbm{1}(u<0)\}$ is the check function, and $K(\cdot)$ represents a kernel function integrating to one. While the above examples of loss functions provide some concrete settings for the proposed approach, we note that our methodology is generally applicable to more general loss functions that satisfy Definition \ref{def:def1}.

\vspace{-.15in}
\subsection{Case without Measurement Errors}
\label{subsec:noiseless}
In this section, we propose an estimator for the Lipschitz continuous loss function when predictors are observed without noise, which is subsequently generalized to measurement error models (that is our focus) in Section \ref{subsec:noisy}. This setting is similar to \cite{dedieu2019sparse}, but our methodology based on feasible sets is distinct from the $L_1$ constrained regularization adopted in that article, and the corresponding theoretical derivations and implementations are also distinct. Our treatment assumes that the noiseless predictors are normally distributed, i.e. $\bm{x}_i\sim N(\bm{0},\Sigma_x),i=1,\ldots,n$. Further, we denote the gradient of the empirical loss $\mathcal{L}(\bm{\beta})$ as $S(\bm{\beta})$ where
$ S(\bm{\beta}) = (1/n)\sum_{i=1}^n\partial f(\langle \bm{x}_i,\bm{\beta}\rangle; y_i)\bm{x}_i$.
From the definition of the true coefficient $\bm{\beta}^*$ we know that the expectation of its gradient function is equal to $\bm{0}$. Therefore intuitively we expect the gradient $S(\bm{\beta}^*)$ to be bounded within a neighborhood of zero with probability tending to one. This motivates us to define the feasible set $\mathbb{C}(\lambda)$ based on the gradient as $ \mathbb{C}(\lambda) = \{\bm{\beta}\in \mathbb{R}^p:\|S(\bm{\beta})\|_\infty\le \lambda \} $, where $\lambda$ acts as a threshold and $||\cdot ||_\infty$ denotes the supremum norm. The feasible set contains a set of all admissible solutions for the proposed estimator, and should be justified as long as $\bm{\beta}^*$ belongs to this set with high probability. This is indeed the case for suitable choice of $\lambda$, as confirmed in the following result.

\begin{lemma}
Let $\lambda = \sqrt{\frac{\phi\log p}{n}}$ and $\sigma_x^2=\|\Sigma_x\|_{op}$ denotes the spectral norm of matrix $\Sigma_x$, then $\bm{\beta}^*\in\mathbb{C}(\lambda)$ with probability at least $[1-2p^{1-(\phi/2L^2\sigma_x^2)}]$, where $\phi>2L^2\sigma_x^2$.
\label{lem:lem1.1}
\end{lemma}

Under our high-dimensional settings $\log(p)/n$ becomes negligible, or equivalently $\lambda$ approaches zero, for increasing $n$ and $p$. Hence, Lemma \ref{lem:lem1.1} suggests that the feasible set only contains those $\bm{\beta}$ values that encourage  $S(\bm{\beta})$ to become increasingly close to zero, at least for large $n$. In addition to the proposed estimator belonging to $\mathbb{C}(\lambda)$, another key consideration is sparsity. In other words, it is desirable to define an estimator with the highest sparsity level, as measured by $L_1$ norm (represented as $||\cdot||_1$). Such an estimator will prevent the solution from admitting excessive false positive signals and to replicate the behavior of  $\bm{\beta}^*$ that is assumed to be sparse. Following this argument, we define our estimator as:
\begin{equation}
\hat{\bm{\beta}} = \argmin_{\bm{\beta}\in\mathbb{C}(\lambda)}\|\bm{\beta}\|_1. \label{eq:estimatorL1}
\end{equation}
 
The proposed estimator in (\ref{eq:estimatorL1}) can be viewed as a generalization of the Dantzig selector \citep{candes2007dantzig} for linear regression to the case with Lipschitz continuous losses. On closer inspection of the constraint of the Dantzig selector, we can see that it is exactly the gradient function of the least squares loss of a linear regression model. We also note that the proposed estimator in (\ref{eq:estimatorL1}) is fundamentally different from related work based on $L_1$ constrained loss functions in \citep{dedieu2019sparse}. In particular, that approach minimizes an optimization criteria combining the Lipschitz loss and an $L_1$ penalty, while constraining the solution to lie within a ball of pre-specified radius. In contrast, the proposed approach minimizes the $L_1$ norm of the coefficient vector that is constrained to lie within a feasible set characterized by bounded empirical gradient. Given the construction of the proposed estimator in (\ref{eq:estimatorL1}) that requires it to have the smallest $L_1$ norm among the class of all estimators belonging to $\mathbb{C}(\lambda)$ and further given the fact that $\bm{\beta}^*\in \mathbb{C}(\lambda)$ is bounded, it is evident that $||\hat{\bm{\beta}}||_1$ is bounded. Hence one need not  project the solution onto a ball of pre-specified radius as in \cite{dedieu2019sparse}. Therefore, our treatment using feasible sets provides a more intuitive approach based on the behavior of gradients, and can be considered non-trivial generalizations of methods in \cite{rosenbaum2010sparse, rosenbaum2013improved} that were originally designed for linear regression approaches involving high-dimensional covariates with measurement errors.

Any desirable estimator $\hat{\bm{\beta}}$ can be expected to lie close to the true parameter $\bm{\beta}^*$, such that the difference $\hat{\bm{h}} = \hat{\bm{\beta}} - \bm{\beta}^*$ is not only bounded but increasingly small. It turns out that the difference $\hat{\bm{h}}$ can be proved to lie in a cone set $\mathcal{H}$ (see Lemma \ref{lem:lem1.2}). The cone set condition ensures that the norm of the estimated coefficients corresponding to truly zero effects ($\beta^*_j=0$) is bounded above by the difference between the estimated and true non-zero coefficients. The concept of cone set is also crucial in defining the restricted strong convexity (RSC) condition, which constitutes an important pillar for ensuring reasonable behavior of the estimators in the absence of strong convexity in high-dimensional settings \citep{ma2022multi,negahban2012unified}. The RSC condition  guarantees that the loss function is not too flat in some restricted sets such that a closeness in the values of the loss function corresponding to ${\bm \beta}^*$ and $\hat{\bm \beta}$ translates to tight error bounds. The following Lemma and definition capture the cone set result and the RSC condition, respectively.

\begin{lemma}
Assume $\bm{\beta}^*$ is $k$-sparse. For any $\bm{\beta}\in\mathbb{R}^p$ and satisfying that $\|\bm{\beta}\|_1\le\|\bm{\beta}^*\|_1$, the difference $\bm{h} = \bm{\beta}-\bm{\beta}^*$ lies in a cone set $\mathcal{H} := \{\bm{h}\in\mathbb{R}^p:\|\bm{h}_{s^c}\|_1\le \|\bm{h}_s\|_1\}$, where set $s = \{j\in\{1,\cdots,p\}: \beta^*_j\ne 0\}$, and complement set $s^c = \{j\in\{1,\cdots,p\}:\beta^*_j=0\}$. 
\label{lem:lem1.2}
\end{lemma}

\begin{definition}
\label{def:rsc1}
(\textbf{RSC condition 1 - without measurement errors}) There exists $\tau>0$ such that $\mathcal{L}(\bm{\beta}^*+\bm{h}) - \mathcal{L}(\bm{\beta}^*) - \langle S(\bm{\beta}^*),\bm{h}\rangle \ge \tau\|\bm{h}\|_2^2$ for all $\bm{h}\in\mathcal{H}$ and $\|\bm{h}\|_1\le 2\|\bm{\beta}^*\|_1$.
\end{definition}

With the theoretical results from \cite{raskutti2010restricted}, we can prove this RSC condition to hold with high probability under certain assumptions, as detailed in Lemma \ref{lem:lem1.3}. Throughout the article, we denote $\Sigma^{1/2} = \sqrt{\Sigma}$ as the square root of a positive semi-definite matrix $\Sigma$ that satisfies the conditions: (i) $\Sigma^{1/2}$ is positive semi-definite; and (ii) $\Sigma^{1/2}\Sigma^{1/2} = \Sigma$. Further for $Var({\bf x})=\Sigma_x$, we define $\kappa_1=(1/4)\lambda_{\min}(\Sigma_x^{1/2})$ and $\kappa_2=9\sqrt{\max_{j=1,\cdots,p}(\Sigma_x)_{jj}}$.

\begin{lemma}
Assume $\min_{\|\bm{h}\|_1\le 2\|\bm{\beta}^*\|_1}|\partial^2 f(\langle\bm{x}_i,\bm{\beta}^*+\bm{h}\rangle;y_i)|\ge M_1>0$ for all $({\bf x}_i,y_i)\in \mathcal{D}_n$. If  $n>4(\kappa_2/\kappa_1)^2 k\log p$, then the RSC condition 1 holds with $\tau = \frac{M_1}{2}\Big(\kappa_1 - 2\kappa_2\sqrt{\frac{k\log p}{n}}\Big)^2$ and with probability at least $(1-c_1\exp(-c_2 n))$ for positive constants $c_1$ and $c_2$.
\label{lem:lem1.3}
\end{lemma}

For the RSC condition 1 to hold in Lemma \ref{lem:lem1.3}, the ratio $\log(p)/n$ needs to be bounded by $\big[4(\kappa_2/\kappa_1)^2 k \big]^{-1}$ that depends on the true sparsity level $k$ and the ratio $\kappa_2/\kappa_1$ that is related to the condition number encountered in random matrix theory literature \citep{edelman1988eigenvalues}. The above Lemma requires the assumption  $\min_{\|\bm{h}\|_1\le 2\|\bm{\beta}^*\|_1}|\partial^2 f(\langle\bm{x}_i,\bm{\beta}^*+\bm{h}\rangle;y_i)|\ge M_1>0$, which guarantees a strictly positive lower bound on the second derivative with respect to the inner product $\langle\bm{x}_i,\bm{\beta}^*+\bm{h}\rangle$. Here, we also note that $\frac{1}{n}\sum_{i=1}^n \partial^2 f(\langle \bm{x}_i, \bm{\beta} \rangle; y_i)$ is different from the derivative of the gradient function that is given as $ \frac{\partial S(\bm{\beta})}{\partial\bm{\beta}} = \frac{1}{n}\sum_{i=1}^n[\partial^2 f(\langle\bm{x}_i,\bm{\beta}\rangle;y_i)]\bm{x}_i\bm{x}_i^T$. The latter can be interpreted as the equivalent of the Hessian matrix in linear regression. For $p<n$ cases, $\sum \bm{x}_i\bm{x}_i^T$ is guaranteed to be positive semi-definite that results in strong convexity, but this does not hold in scenarios when $p>>n$. In such scenarios, the RSC condition is required in lieu of strong convexity, to guarantee reasonable estimators. In order for the RSC condition to hold, our approach requires the additional assumption $\min_{\|\bm{h}\|_1\le 2\|\bm{\beta}^*\|_1}|\partial^2 f(\langle\bm{x}_i,\bm{\beta}^*+\bm{h}\rangle;y_i)|\ge M_1>0$ that can be shown to hold with high probability for several routinely used loss functions. This is illustrated in the following Lemma.

\begin{lemma}
\label{lem:lem1.4}
The assumption stated in Lemma \ref{lem:lem1.3} that $\min_{\|\bm{h}\|_1\le 2\|\bm{\beta}^*\|_1}|\partial^2 f(\langle\bm{x}_i,\bm{\beta}^*+\bm{h}\rangle;y_i)|\ge M_1>0$ for all $({\bf x}_i,y_i)\in \mathcal{D}_n$ holds with high probability tending to one for the special cases involving logistic loss, smooth hinge loss and conquer loss.
\end{lemma}

Given the above definitions, we are now in a position to state our main error bound results. Theorem \ref{theo:theo1.1} provides explicit finite sample error bounds for the proposed estimator.

\begin{theorem}
Assume $\bm{\beta}^*$ is $k$-sparse and $\min_{\|\bm{h}\|_1\le 2\|\bm{\beta}^*\|_1}|\partial^2 f(\langle\bm{x}_i,\bm{\beta}^*+\bm{h}\rangle;y_i)|\ge M_1>0$ for all $({\bf x}_i,y_i)\in \mathcal{D}_n$. Then with probability at least $[1-2p^{1-(\phi/2L^2\sigma_x^2)}-c_1\exp(-c_2 n)]$, the estimator in (\ref{eq:estimatorL1}) satisfies the error bounds:
$ \|\hat{\bm{\beta}}-\bm{\beta}^*\|_2 \le \frac{4\sqrt{k}}{\tau}\sqrt{\frac{\phi\log p}{n}} $, and $ \|\hat{\bm{\beta}}-\bm{\beta}^*\|_1 \le \frac{8k}{\tau}\sqrt{\frac{\phi\log p}{n}} $, where $\phi>2L^2\sigma_x^2$, $L$ is Lipschitz constant and constants $c_1,c_2>0,$.
\label{theo:theo1.1}
\end{theorem}

\begin{remark}
From Theorem \ref{theo:theo1.1}, we can see that as long as $k^2\log p = o(n)$ where $o(\cdot)$ denotes the small-o notation, then both error bounds would vanish to zero as $n$ and $p$ increase to infinity. This requires that the number of truly nonzero coefficients remains relatively small that is routinely assumed in literature, while the total number of coefficients $p$ is allowed to increase with $n$ exponentially, reflecting high-dimensional cases.
\end{remark}

Although Theorem \ref{theo:theo1.1} ensures good estimation performance in terms of finite sample error bounds, the properties of the estimator in terms of feature selection are not immediately clear. Fortunately, one can threshold the proposed estimator in a manner that enables accurate feature selection by guaranteeing sign consistency. The next result in Theorem \ref{theo:theo1.2} establishes the sign consistency, which illustrates that the threshold estimator is able to accurately identify the truly non-zero coefficients as long they are not exceedingly small. Further, the threshold depends on  $k\sqrt{\frac{\phi\log p}{n}}$ and hence can be decreased with increasing sample sizes. Taken together, the results in Theorems \ref{theo:theo1.1}-\ref{theo:theo1.2} establish desirable finite sample operating characteristics for a general class of Lipschitz continuous loss functions.

\begin{theorem}
(\textbf{Sign Consistency}) Let $\tau_1 = \frac{8k}{\tau}\sqrt{\frac{\phi\log p}{n}}$ and define a threshold version of estimator as $\Tilde{\beta}_j = \hat{\beta}_j{\bf 1}\{|\hat{\beta}_j|>\tau_1\},\mbox{ } j=1,\cdots,p$.
If the error bounds in Theorem \ref{theo:theo1.1} hold and we have $\min_{j\in s}|\beta_j^*|>2\tau_1$, then $sign \Tilde{\beta}_j = sign \beta_j^*$, with high probability.
\label{theo:theo1.2}
\end{theorem}

\vspace{-.15in}
\subsection{Case with Measurement Errors}
\label{subsec:noisy}
We now extend the proposed method in the presence of measurement error on high-dimensional covariates. In such settings, the true coefficient is not guaranteed to be within the feasible set $\mathbb{C}(\lambda)$ defined in Section \ref{subsec:noiseless} with high probability. See Section 7 in \cite{rosenbaum2010sparse} for more detailed explanations for the linear regression case. Hence, the feasible set needs to be modified in a suitable manner in the presence of measurement error, such that it will ensure desirable finite sample error bounds on the proposed estimators.

We still assume the true predictors $\bm{x}\sim N(\bm{0},\Sigma_x)$ as in Section \ref{subsec:noiseless}. However in the scenario with measurement errors, one observes a contaminated version of the true predictor as $\bm{w}_i$'s where
$\bm{w}_i = \bm{x}_i + \bm{u}_i, \mbox{ } i=1,\cdots,n. $. The $\bm{u}_i$'s are the measurement errors, which are assumed to independently follow $N(\bm{0},\Sigma_u)$. The $\bm{x}_i$'s and $\bm{u}_i$'s are assumed to be independent from each other. We denote the noisy dataset as $\mathcal{D}^w_n:= \{(\bm{w}_i,y_i)\in\mathbb{R}^p\times\mathcal{Y}, i=1,\cdots,n\}$.

The above constructions for the measurement error are routinely assumed in literature involving linear regression models \citep{rosenbaum2010sparse,loh2012high,ma2022multi}. Based on the new design matrix, the corresponding empirical loss and gradient functions would be
$ \mathcal{L}_w(\bm{\beta}) = (1/n)\sum_{i=1}^n f(\langle \bm{w}_i,\bm{\beta}\rangle; y_i) $ and
$ S_w(\bm{\beta}) = (1/n)\sum_{i=1}^n\partial f(\langle \bm{w}_i,\bm{\beta}\rangle; y_i)\bm{w}_i $. Let $O$ denote the big-O notation. We make the following assumptions.

{\noindent \it (A1)} $n\sigma_u^2=O(1)$ where $\sigma_u^2=\|\Sigma_u\|_{op}$, and $||\cdot||_{op}$ denotes the operator norm.

{\noindent \it (A2)} Without loss of generality, we will assume the design matrix $W=(\bm{w}_1,\cdots,\bm{w}_n)^T$ has been standardized such that all the diagonal elements in $\frac{W^TW}{n}$ are equal to 1.

{\noindent \it (A3)} The true coefficient $\bm{\beta}^*$ satisfies  $\bm{\beta}^* = \argmin\{E(f(\langle \bm{x},\bm{\beta}\rangle; y))\}$ and it is $k$-sparse.

We note that it is not unreasonable to have assumption {\it (A1)} for practical applications. In medical imaging analysis, as we collect the image data at finer and finer resolutions, which means more voxels and larger $p$, we expect the scale of measurement errors to go down. This implies that $\sigma_u^2$ is inversely proportional to the number of voxels $p$, i.e. $\sigma^2_u= O(1/p)$.  Typically the number of samples in imaging studies is far less compared to the voxel numbers ($p>>n$). Thus the assumption $n\sigma_u^2=O(1)$ will be applicable in a general sense for high-dimensional cases. Assumption {\it (A2)} is a common argument in the literature \citep{fan2011nonconcave, rosenbaum2010sparse}. Finally, Assumption {\it (A3)} captures the relationship between the true coefficient $\bm{\beta}^*$ and the outcome $y\in\mathcal{Y}$ via the loss function that depends on the true (noiseless) predictors $\bm{x}$, and it is similar to the noiseless scenario in Section \ref{subsec:noiseless}.

In order to construct the feasible set for the measurement error case, we enlarge the feasible band for $\|S_w(\bm{\beta})\|_\infty$ by allowing the boundary of the feasible set to adapt to the $L_1$ norm of $\bm{\beta}$, which is similar to the approach in \citep{rosenbaum2010sparse} for linear regression models. The modified feasible set and the estimator are defined below:
\begin{eqnarray}
\mathbb{C}^w(\lambda,\gamma) = \{\bm{\beta}\in \mathbb{R}^p:\|S_w(\bm{\beta})\|_\infty\le \lambda + \gamma\|\bm{\beta}\|_1\}, \quad \hat{\bm{\beta}}^w = \argmin_{\bm{\beta}\in\mathbb{C}^w(\lambda,\gamma)}\|\bm{\beta}\|_1. \label{eq:MUL-noisy}
\end{eqnarray}
The modified feasible set with expanded boundaries helps the estimator to compensate for the errors in the gradient function $S_w(\bm{\beta})$ that arise due to the noisy covariates. In particular, while the empirical gradient function $S(\bm{\beta})$ corresponding to the true(noiseless) predictors is expected to be close to zero as per arguments in Section \ref{subsec:noiseless}, the same can not be guaranteed for $S_w(\bm{\beta})$ computed in the presence of measurement errors. Further, the deviations for this gradient function from zero may potentially increase with the number of non-zero elements in $\bm{\beta}$. The construction of the feasible set $\mathbb{C}^w(\lambda,\gamma)$ accommodates this feature by expanding the boundaries of the set in a manner that is proportional to the norm $||\bm{\beta}||_1$. It also guarantees that $\bm{\beta}^*$ will belong to $\mathbb{C}^w(\lambda,\gamma)$ with high probability for properly chosen parameters $\lambda$ and $\gamma$, which is a key requirement. This result is captured in the following Lemma \ref{lem:lem2.1}.

\begin{lemma}
Let $\lambda = \sqrt{\frac{\phi_1\log p}{n}}$ and $\gamma = M_2\sqrt{\frac{\phi_2\log n}{n}}$ where $M_2 = \max|\partial^2 f(\cdot;y)|$. Then under Assumptions (A1)-(A3), $\bm{\beta}^*\in\mathbb{C}^w(\lambda,\gamma)$ with probability at least $\{1-2p^{1-[\phi_1/2L^2(\sigma_x^2+\sigma_u^2)]}-2n^{(1-\phi_2/2n\sigma_u^2)}\}$,  where $\phi_1>2L^2(\sigma_x^2+\sigma_u^2)$ and $\phi_2>2n\sigma_u^2$.
\label{lem:lem2.1}
\end{lemma}

The second order derivative in Lemma \ref{lem:lem2.1} exists using Definition \ref{def:def1}, and it is usually bounded for our problems of interest. For example, it is possible to  find a finite upper bound of $M_2$ for the class of Lipschitz continuous losses we consider. This is evident given that $|\partial^2 f(\cdot;y)| = e^t/(1+e^t)^2 \le 1/4$ for logistic loss, and $|\partial^2 f(\cdot;y)| = \frac{1}{2}\sigma^2/[(1-yt)^2+\sigma^2]^{3/2} \le 1/(2\sigma)$ for smooth hinge loss. Further, the parameters $\phi_1$ and $\phi_2$ are constants that can be selected properly to ensure Lemma \ref{lem:lem2.1} holds with high probability.

We note turn our attention to the cone set for the case with measurement errors, which is defined similarly as in Lemma \ref{lem:lem1.2}. It can be shown that the difference $\hat{\bm{h}}^w = \hat{\bm{\beta}}^w-\bm{\beta}^*\in \mathbb{C}^w(\lambda,\gamma)$ lies in the cone set $\mathcal{H} := \{\bm{h}^w\in\mathbb{R}^p:\|\bm{h}^w_{s^c}\|_1\le \|\bm{h}^w_s\|_1\}$.  The corresponding RSC condition when the measurement errors are present is given below.

\begin{definition}
\label{def:rsc2}
(\textbf{RSC condition 2 - with measurement errors}) There exists $\tau_w>0$ such that $ \mathcal{L}_w(\bm{\beta}^*+\bm{h}) - \mathcal{L}_w(\bm{\beta}^*) - \langle S_w(\bm{\beta}^*),\bm{h}\rangle \ge \tau_w\|\bm{h}\|_2^2$ for all $\bm{h}\in\mathcal{H}$ and $\|\bm{h}\|_1\le 2\|\bm{\beta}^*\|_1$.
\end{definition}
 
The curvature $\tau_w$ for the RSC condition 2 with measurement errors is defined similarly as in Lemma \ref{lem:lem1.3}, but now depends on the covariance matrix of the noisy covariates $\Sigma_w=\Sigma_x + \Sigma_u$ via corresponding minimum and maximum eigen values $\kappa_1^w=(1/4)\lambda_{\min}(\sqrt{\Sigma_w})$ and $\kappa_2^w=9\sqrt{\max_{j=1,\cdots,p}(\Sigma_w)_{jj}}$. Moreover, the tolerance in the RSC condition, i.e. $\langle S_w(\bm{\beta}^*),\bm{h}\rangle$, now depends on the observed noisy covariates. It can be shown that the RSC condition 2 holds with high probability in the following Lemma.

\begin{lemma}
Assume that $\min_{\|\bm{h}\|\le 2\|\bm{\beta}^*\|_1}|\partial^2 f(\langle\bm{w}_i,\bm{\beta}^*+\bm{h}\rangle;y_i)|\ge M_1^w>0$ for all $(\bm{w}_i, y_i)\in \mathcal{D}^w_n$. If $n>4(\kappa_2^w/\kappa_1^w)^2 k\log p$, then the RSC condition 2 holds with $\tau_w = \frac{M_1^w}{2}\Big(\kappa_1^w - 2\kappa_2^w\sqrt{\frac{k\log p}{n}}\Big)^2$ with probability at least $(1-c_1^\prime\exp(-c_2^\prime n))$ for positive constants $c_1^\prime$ and $c_2^\prime$.
\label{lem:lem2.2}
\end{lemma}

\begin{remark}
Similar to the results in Lemma \ref{lem:lem1.4}, the assumption in Lemma \ref{lem:lem2.2} that $\min_{\|\bm{h}\|\le 2\|\bm{\beta}^*\|_1}|\partial^2 f(\langle\bm{w}_i,\bm{\beta}^*+\bm{h}\rangle;y_i)|\ge M_1^w>0$ will hold with high probability for the three example loss functions we consider, i.e. logistic loss, smooth hinge loss, and conquer loss.
\end{remark}

Given the above definitions, we can now obtain the error bounds for the estimator in (\ref{eq:MUL-noisy}).

\begin{theorem}
Assume $\min_{\|\bm{h}\|\le 2\|\bm{\beta}^*\|_1}|\partial^2 f(\langle\bm{w}_i,\bm{\beta}^*+\bm{h}\rangle;y_i)|\ge M_1^w>0$ for all $(\bm{w}_i, y_i)\in \mathcal{D}^w_n$ and $n>4(\kappa_2^w/\kappa_1^w)^2 k\log p$. Under (A1)--(A3), we have $ \|\hat{\bm{\beta}}^w-\bm{\beta}^*\|_2 \le \frac{4\sqrt{k}}{\tau_w}\big(\sqrt{\frac{\phi_1\log p}{n}} + M_2\sqrt{\frac{\phi_2\log n}{n}}\|\bm{\beta}^*\|_1\big) $ and $ \|\hat{\bm{\beta}}^w-\bm{\beta}^*\|_1 \le 2\sqrt{k}\|\hat{\bm{\beta}}^w-\bm{\beta}^*\|_2 $ with probability at least $\{1-c_1^\prime\exp(-c_2^\prime n)-2p^{1-[\phi_1/2L^2(\sigma_x^2+\sigma_u^2)]}-2n^{(1-\phi_2/2n\sigma_u^2)}\}$ for $c_1^\prime>0, c_2^\prime>0$, $\phi_1>2L^2(\sigma_x^2+\sigma_u^2)$, $\phi_2>2n\sigma_u^2$.
\label{theo:theo2.1}
\end{theorem}

Compared to the case without noise, the error bound is greater for the case with measurement error, by a term that depends on $||\bm{\beta}^*||_1$. Given that we need $\phi_2>2n\sigma_u^2$ for Theorem \ref{theo:theo2.1} to hold with high probability, the second term may not become negligible even when $n$ is large. We note that an inflated error bound is expected for the case involving noisy covariates, given that the true covariates are not observed. Fortunately, the second term can be shown to be vanishingly small in certain special cases, such as when $\phi_2 =  2n\sigma_u^2\times \tilde{\phi}_2$ for some constant $\tilde{\phi}_2>1$. In this case, the second term in the $L_1$ error bound becomes $\frac{8k}{\tau_w}M_2\sqrt{\frac{\phi_2\log n}{n}} \|\bm{\beta}^*\|_1= \frac{8k}{\tau_w}M_2\sqrt{2\log(n)\tilde{\phi}_2\sigma^2_u}\times \|\bm{\beta}^*\|_1$. As long as $k^2\log(n)\sigma^2_u= o(1)$, the second term (that contributes to an inflated error bound) can be made to vanishingly small.  Therefore in such a special limiting case, as $n$ increases and $\sigma_u$ becomes negligible, the statistical error in the presence of measurement error will start to resemble the error under the setting without measurement errors. Our result  illustrates that the estimator in the presence of measurement error will become essentially as good as the estimator for the case without measurement error for large sample sizes, as the noise level decreases. Although the main focus of the results in Theorem \ref{theo:theo2.1} is in terms of finite sample error bounds, the above arguments provide an insight into the behavior of the estimator in large $n$ settings.

As in the case without measurement error, it is possible to construct a thresholded estimator that satisfies sign consistency for the scenario involving noisy features. Similar to that case, the threshold can be relaxed to smaller values for increasing $n$ when we have greater confidence on the parameter estimates as defined by tighter error bounds. This is captured in the following Theorem. 

\begin{theorem}
(\textbf{Sign Consistency}) Let $\tau_1^w = \frac{8k}{\tau_w}\Big(\sqrt{\frac{\phi_1\log p}{n}} + R^*M_2\sqrt{\frac{\phi_2\log n}{n}}\Big)$ where $R^*\ge\|\bm{\beta}^*\|_1$ and define a threshold version of estimator as $\Tilde{\beta}_j^w = \hat{\beta}_j^w{\bf 1}\{|\hat{\beta}_j^w|>\tau_1^w\},\mbox{ } j=1,\cdots,p$. 
If the error bounds in Theorem \ref{theo:theo2.1} hold and $\min_{j\in s}|\beta_j^*|>2\tau_1^w$, then $sign \Tilde{\beta}_j^w = sign \beta_j^*$. \label{theo:theo2.2}
\end{theorem}

\vspace{-.15in}
\subsection{Extension to Lasso Analog}
\label{subsec:analog}
Although the estimators defined in Sections \ref{subsec:noiseless}-\ref{subsec:noisy} are desirable in terms of having attractive finite sample error bounds and feature selection properties via sign consistency, it may potentially run into computational bottlenecks for high dimensions due to the linear programming involved in implementing the method in practice (see Section \ref{sec:compute}). Ideally, one would like the proposed approach to computationally scale up to a large number of covariates. To this end, we propose a computationally scalable Lasso-type analog estimator for our problem that is related to the original estimators in a manner similar  to the connection between the Lasso and the Dantzig selector  \citep{rosenbaum2010sparse}. For our cases of interest involving measurement errors, the Lasso analog criteria takes the form:
\begin{equation}
\min_{||\bm{\beta}||_1\le 2\|\bm{\beta}^*\|_1}\{\mathcal{L}_w(\bm{\beta}) + \lambda_2\|\bm{\beta}\|_1 + (\gamma_2/2)\|\bm{\beta}\|_1^2 \}.
\label{eqn:analog2}
\end{equation}
We note that the above estimator reduces to the penalized problem: $\min_{||\bm{\beta}||_1\le 2\|\bm{\beta}^*\|_1} \{\mathcal{L}(\bm{\beta}) + \lambda_2\|\bm{\beta}\|_1\}$ in the absence of noise that was considered in \cite{dedieu2019sparse}, and can be considered as a special case under our set-up. Further we note that the Lasso analog estimator was briefly discussed in \cite{rosenbaum2010sparse} for linear regression models involving noisy covariates, but without additional details on theoretical and practical implications.

To appreciate why there is a similarity between (\ref{eqn:analog2}) and (\ref{eq:MUL-noisy}), note that for $\bm{\beta}$ to achieve the minimum of such a convex criterion (\ref{eqn:analog2}), it is necessary and sufficient to satisfy the KKT conditions: (i) $-[S_w(\bm{\beta})]_j = (\lambda_2 + \gamma_2\|\bm{\beta}\|_1) sign(\beta_j) \mbox{ if } \mbox{ }\beta_j\ne 0;$ and (ii) $|[S_w(\bm{\beta})]_j| \le \lambda_2 + \gamma_2\|\bm{\beta}\|_1, \mbox{ if } \mbox{ } \beta_j= 0$. It is straightforward to see that the optimizers satisfying the KKT conditions are tightly contained in the modified feasible set $\mathbb{C}^w(\lambda_2,\gamma_2)$. More importantly, based on the results from Lemma \ref{lem:analog_cone}, the true coefficients $\bm{\beta}^*$ can be shown to lie in a similarly modified feasible set $\mathbb{C}^w(\lambda_2/2,\gamma_2/2)$ with high probability, which serves to validate the proposed estimator. Another powerful validation comes from the fact that the difference between the lasso analog estimator and the true coefficients can be shown to lie within a well-behaved cone set $\mathcal{H}_L$, which is a crucial requirement for establishing finite sample error bounds in the sequel. These results are captured by the following Lemma \ref{lem:analog_cone}.

\begin{lemma}
Let $\lambda_2=2\sqrt{\frac{\phi_1\log p}{n}}$ and $\gamma_2=2M_2\sqrt{\frac{\phi_2\log n}{n}}$, then $\bm{\beta}^*\in\mathbb{C}^w(\lambda_2/2,\gamma_2/2)$ with probability at least $\{1-2p^{1-[\phi_1/2L^2(\sigma_x^2+\sigma_u^2)]}-2n^{(1-\phi_2/2n\sigma_u^2)}\}$ where $\phi_1>2L^2(\sigma_x^2+\sigma_u^2)$ and $\phi_2>2n\sigma_u^2$. Further under assumptions (A1)-(A3), the difference $\hat{\bm{h}}_L^w = \hat{\bm{\beta}}_L^w - \bm{\beta}^*$ belongs to the cone set $\mathcal{H}_L:=\{\bm{h}\in\mathbb{R}^p:\|\bm{h}_{s^c}\|_1\le 3\|\bm{h}_s\|_1\}$, where $\hat{\bm{\beta}}_L^w$ denotes the optimizer in (\ref{eqn:analog2}).
\label{lem:analog_cone}
\end{lemma}

Further, we can also define an RSC condition below on the cone set $\mathcal{H}_L$ and show it to hold with high probability, which is elaborated in the following Lemma \ref{lem:analog_rsc}.

\begin{definition}
(\textbf{RSC condition 3}) There exists $\tau_L^w>0$ such that $\mathcal{L}_w(\bm{\beta}^*+\bm{h}) - \mathcal{L}_w(\bm{\beta}^*) - \langle S_w(\bm{\beta}^*),\bm{h}\rangle \ge \tau_L^w\|\bm{h}\|_2^2$ for all $\bm{h}\in\mathcal{H}_L$ and $\|\bm{h}\|_1\le 3\|\bm{\beta}^*\|_1$.
\end{definition}

\begin{lemma}
\label{lem:analog_rsc}
Assume that $\min_{\|\bm{h}\|_1\le 3\|\bm{\beta}^*\|_1}|\partial^2 f(\langle\bm{w}_i,\bm{\beta}^*+\bm{h}\rangle;y_i)|\ge M_3^w>0$ for all $(\bm{w}_i,y_i)\in\mathcal{D}_n^w$. If $n>16(\kappa_2^w/\kappa_1^w)^2 k\log p$ where $\kappa_1^w$ and $\kappa_2^w$ are defined as in Lemma \ref{lem:lem2.2}, then the RSC condition 3 holds for $\tau_L^w = \frac{M_3^w}{2}\Big(\kappa_1^w - 4\kappa_2^w\sqrt{\frac{k\log p}{n}}\Big)^2$ with probability at least $(1-c_1^\prime\exp(-c_2^\prime n))$ for positive constants $c_1^\prime$ and $c_2^\prime$.
\end{lemma}

\begin{remark}
\label{remark:analog_rsc}
The assumption $\min_{\|\bm{h}\|_1\le 3\|\bm{\beta}^*\|_1}|\partial^2 f(\langle\bm{w}_i,\bm{\beta}^*+\bm{h}\rangle;y_i)|\ge M_3^w>0$ for all $(\bm{w}_i,y_i)\in\mathcal{D}_n^w$ can be shown to hold with high probability under the three loss functions of interest, similar to Lemma \ref{lem:lem1.4}.
Moreover, compared to estimators in (\ref{eq:MUL-noisy}), the Lasso analog estimator requires a higher sample size for the RSC condition to hold with high probability.  
\end{remark}

Similar to the estimators based on feasible sets in Section \ref{subsec:noisy}, it is possible to derive finite sample error bounds under the Lasso analog as established below in Theorem \ref{theo:analog_bound}. Although the order of the error bound is similar compared to Theorem \ref{theo:theo2.1}, the scaling constants are slightly inflated under the Lasso analog estimator. This is not unexpected given that it can be considered an approximation of the original estimators in (\ref{eq:MUL-noisy}). Such an inflation factor may be viewed as the price one must pay in order to make the approach more computationally scalable to higher dimensions.

\begin{theorem}
\label{theo:analog_bound}
Assume $\min_{\|\bm{h}\|_1\le 3\|\bm{\beta}^*\|_1}|\partial^2 f(\langle\bm{w}_i,\bm{\beta}^*+\bm{h}\rangle;y_i)|\ge M_3^w>0$ for all $(\bm{w}_i,y_i)\in\mathcal{D}_n^w$ and $n>16(\kappa_2^w/\kappa_1^w)^2 k\log p$. Under Assumptions (A1)-(A3), the error bounds under the Lasso analog are given as: $ \|\hat{\bm{\beta}}_L^w-\bm{\beta}^*\|_2 \le \frac{12\sqrt{k}}{\tau_L^w}\bigg[\sqrt{\frac{\phi_1\log p}{n}} + M_2\sqrt{\frac{\phi_2\log n}{n}}\|\bm{\beta}^*\|_1\bigg] $ and $\|\hat{\bm{\beta}}_L^w-\bm{\beta}^*\|_1 \le 4\sqrt{k} \|\hat{\bm{\beta}}_L^w-\bm{\beta}^*\|_2 $ with probability at least $\{1-2p^{1-[\phi_1/2L^2(\sigma_x^2+\sigma_u^2)]}-2n^{(1-\phi_2/2n\sigma_u^2)}-c_1^\prime\exp(-c_2^\prime n)\}$ for constants $c_1^\prime>0, c_2^\prime>0$ and $\phi_1>2L^2(\sigma_x^2+\sigma_u^2)$ and $\phi_2>2n\sigma_u^2$.
\end{theorem}

As in previous sections, it is also possible to obtain the sign consistency results for the threshold version of the Lasso analog, as shown in the following Theorem \ref{theo:analog_threshold}.

\begin{theorem}
\label{theo:analog_threshold}
(\textbf{Sign Consistency}) 
Let $\tau_2^w = \frac{48k}{\tau_L^w}\Big[\sqrt{\frac{\phi_1\log p}{n}} + R^*M_2\sqrt{\frac{\phi_2\log n}{n}}\Big]$ where $R^*\ge\|\bm{\beta}^*\|_1$ and define a threshold version of estimator as $(\Tilde{\beta}_L^w)_j = (\hat{\beta}_L^w)_j{\bf 1}\{|(\hat{\beta}_L^w)_j|>\tau_2^w\},\mbox{ } j=1,\cdots,p$. If Theorem \ref{theo:analog_bound} holds and $\min_{j\in s}|\beta_j^*|>2\tau_2^w$, then $sign (\Tilde{\beta}_L^w)_j = sign \beta_j^*$.
\end{theorem}

\vspace{-.25in}
\section{Computations}
\label{sec:compute}
We first propose a computational algorithm for MULL estimators corresponding to Sections \ref{subsec:noiseless}-\ref{subsec:noisy} based on Newton-Raphson iterative method, and subsequently outline a gradient descent algorithm for the Lasso analog method in Section \ref{subsec:analog}, and discuss parameter tuning.

\vspace{-.15in}
\subsection{Computational Algorithms for MULL Estimators}
We propose to utilize Newton-Raphson type of method with first-order approximation to the gradient function at each iteration. In the case without measurement errors, we assume $\bm{\beta}^{(m)}$ to be our estimate at the $m$-th iteration. Then through first-order Taylor expansion, we can obtain an approximation of the gradient function around $\bm{\beta}^{(m)}$ as
$$ S(\bm{\beta}) \approx S(\bm{\beta}^{(m)}) + \frac{\partial S(\bm{\beta})}{\partial\bm{\beta}}\bigg|_{\bm{\beta}=\bm{\beta}^{(m)}}(\bm{\beta}-\bm{\beta}^{(m)}) = \Sigma^{(m)}\bm{\beta} + \nu^{(m)} $$
where $\Sigma^{(m)} = \frac{\partial S(\bm{\beta})}{\partial\bm{\beta}}\Big|_{\bm{\beta}=\bm{\beta}^{(m)}} = \frac{1}{n}\sum_{i=1}^n[\partial^2 f(\langle\bm{x}_i,\bm{\beta}^{(m)}\rangle;y_i)]\bm{x}_i\bm{x}_i^T$ and $\nu^{(m)} = S(\bm{\beta}^{(m)}) - \Sigma^{(m)}\bm{\beta}^{(m)}$. Using this approximation, the computation turns into the problem of minimizing $\|\bm{\beta}\|_1$ for $\bm{\beta}\in\{\bm{\beta}\in\mathbb{R}^p:\|\Sigma^{(m)}\bm{\beta} + \nu^{(m)}\|_\infty\le\lambda\}$, which translates into a linear programming that can be solved by standard software. We formalize the linear programming problem here:
\begin{equation*}
\begin{split}
&\min_{\bm{b}^+,\bm{b}^-}\mbox{ }  \bm{1}_p^T(\bm{b}^+ + \bm{b}^-) \mbox{ such that } \bm{b}^+\ge\bm{0},\mbox{ }\bm{b}^-\ge\bm{0}, \\
& \Sigma^{(m)}(\bm{b}^+ - \bm{b}^-) \le \lambda\bm{1}_p - \nu^{(m)},\mbox{ }  -\Sigma^{(m)}(\bm{b}^+ - \bm{b}^-) \le \lambda\bm{1}_p + \nu^{(m)}.
\end{split}
\end{equation*}
where $\bm{1}_p$ denotes the vector of length $p$ with entries all equal to 1. We obtain the estimation of $\bm{\beta}$ at iteration $m+1$ as $\bm{\beta}^{(m+1)} = \hat{\bm{b}}^+ - \hat{\bm{b}}^-$. This process is repeated until convergence.

As for the computation with measurement errors when using the modified feasible set $\mathbb{C}^w(\lambda, \gamma)$, we need the approximation of the gradient function $S_w(\bm{\beta})$ around $\bm{\beta}^{(m)}$ such that
$$ S_w(\bm{\beta}) \approx S_w(\bm{\beta}^{(m)}) + \frac{\partial S_w(\bm{\beta})}{\partial\bm{\beta}}\bigg|_{\bm{\beta}=\bm{\beta}^{(m)}}(\bm{\beta}-\bm{\beta}^{(m)}) = \Sigma_w^{(m)}\bm{\beta} + \nu_w^{(m)} $$
where $\Sigma_w^{(m)} = \frac{\partial S_w(\bm{\beta})}{\partial\bm{\beta}}\Big|_{\bm{\beta}=\bm{\beta}^{(m)}} = \frac{1}{n}\sum_{i=1}^n[\partial^2 f(\langle\bm{w}_i,\bm{\beta}^{(m)}\rangle;y_i)]\bm{w}_i\bm{w}_i^T$ and $\nu_w^{(m)} = S_w(\bm{\beta}^{(m)}) - \Sigma_w^{(m)}\bm{\beta}^{(m)}$. Here, the computation translates to minimizing $\|\bm{\beta}\|_1$ for $\bm{\beta}\in\{\bm{\beta}\in\mathbb{R}^p:\|\Sigma_w^{(m)}\bm{\beta} + \nu_w^{(m)}\|_\infty\le\lambda + \gamma\|\bm{\beta}\|_1\}$, which can be written as a linear programming problem:
\begin{equation*}
\begin{split}
&\min_{\bm{b}^+,\bm{b}^-}\mbox{ }  \bm{1}_p^T(\bm{b}^+ + \bm{b}^-) \mbox{ such that } \bm{b}^+\ge\bm{0},\mbox{ }\bm{b}^-\ge\bm{0}, \\
& (\Sigma_w^{(m)}-\gamma J_p)\bm{b}^+ - (\Sigma_w^{(m)}+\gamma J_p)\bm{b}^- \le \lambda\bm{1}_p - \nu_w^{(m)}, \\
& - (\Sigma_w^{(m)}+\gamma J_p)\bm{b}^+ + (\Sigma_w^{(m)}-\gamma J_p)\bm{b}^- \le \lambda\bm{1}_p + \nu_w^{(m)},
\end{split}
\end{equation*}
where $J_p$ denotes the $p$ by $p$ square matrix with entries all equal to $1$.

\vspace{-.15in}
\subsection{Computational Algorithms for Lasso Analog Estimators}

Lasso analog estimation in problem (\ref{eqn:analog2}) can be solved by projected gradient-based optimization algorithm. The estimate at the $t$-th iteration is denoted as $\bm{\beta}^{(t)}$, then we obtain the estimate at iteration $(t+1)$ as $\bm{\beta}^{(t+1)} = \prod_R(\bm{\beta}^{(t)} - \eta g(\bm{\beta}^{(t)}))$ where $\eta>0$ is the stepsize parameter, $g(\bm{\beta}) = S_w(\bm{\beta}) + (\lambda_2 + \gamma_2\|\bm{\beta}\|_1)sign(\bm{\beta})$ and $\prod_R(\cdot)$ denotes the projection operator onto the $L_1$ ball of radius $R$. We use the spectral projected gradient (SPG) algorithm to learn the stepsize with non-monotone line search \citep{birgin2000nonmonotone}. The $L_1$ projection can be computed efficiently using algorithm from \cite{duchi2008efficient}.

\vspace{-.15in}
\subsection{Parameter Tuning and Initialization}
It is critical to tune the model parameters in order for the algorithm to perform optimally in practice. For the parameters $(\lambda,\gamma)$ and $(\lambda_2, \gamma_2)$ involved in the feasible set for the MULL estimator and the Lasso analog respectively, we can conduct cross validation (CV) on a selected grid of candidate values. The candidate values for parameter $\lambda$ and $\lambda_2$ would be around the order of $\sqrt{\log p/n}$, while the candidates for $\gamma$ and $\gamma_2$ would be around the order of $\sqrt{\log n/n}$, as indicated by results in Section \ref{sec:method}. The evaluation criterion for the CV procedure can either be the mis-classification rate on the testing samples for classification problems, or prediction errors in terms of the check loss \citep{wang2012quantile} for quantile regression.

In addition to the above parameters, we also need to tune $\sigma^2$, for the smooth hinge loss function. When $\sigma^2$ is equal to 0, the smooth hinge loss is exactly the original hinge loss. As $\sigma^2$ increases, the corresponding smooth hinge loss function gets further away from the original hinge loss. If $\sigma^2$ becomes too large, it would dominate the loss function that may adversely affect the classification performance. On the other hand, if $\sigma^2$ gets too close to 0 and resembles the original hinge loss, the implementation may become potentially unstable when the inner product $\bm{w}^T\bm{\beta}$ is around zero, which can lead to ill conditioning for the linear programming algorithm in the computation. Given all these considerations, we want to pick a value for $\sigma^2$ which is neither too large nor too small. We tried several choices for $\sigma^2$, and chose $\sigma^2=4$ in our empirical experiments, which provides stable results.

A warm start for the computational algorithm is also helpful for improving the performance under the MULL methods. A good starting point can be the estimate from the Lasso method \citep{tibshirani1996regression}, or the generalized Dantzig selector \citep{candes2007dantzig}. We also implement a thresholding procedure for our MULL estimator and the Lasso analog for feature selection, as suggested by the sign consistency theories. Any coefficients having an estimate below a percentage level of the maximum absolute value of all coefficients will be set to exactly zero. The thresholds may be chosen via cross-validation during implementation.

\vspace{-.25in}
\section{Simulations}
\label{sec:sims}

\vspace{-.15in}
\subsection{Classification Problems}
\label{subsec:sims_class}
We examine the empirical performance of our proposed method in several different scenarios of binary classification problem that is subject to measurement errors, and also compare with other competing methods including logistic regression with $L_1$-norm penalty (`L1logistic'), $L_1$-norm SVM, GDS and GMUS. These competing methods are realized through \texttt{R} packages including \texttt{glmnet} \citep{friedman2010regularization}, \texttt{penalizedSVM} \citep{becker2009penalizedsvm} and \texttt{hdme} \citep{sorensen2018covariate}. We denote our proposed classification approach under the MUL methods as matrix uncertainty classifier (MUC), and denote the MUC corresponding to the logistic and the smooth hinge loss as `muc.logistic' and `muc.smhinge' respectively. We denote the corresponding Lasso analog methods as `analog.logistic' and `analog.smhinge' respectively. We consider three distinct data generation schemes, which are presented below.

{\em Scheme 1}: $P(y=1) = P(y=0) = 0.5$, $\bm{x}|(y=1) \sim \mathrm{N}_p(\bm{\mu}, \bm{\Sigma})$, $\bm{x}|(y=0) \sim \mathrm{N}_p(-\bm{\mu}, \bm{\Sigma})$, $\bm{\mu} = (0.1, 0.2, 0.3, 0.4, 0.5, 0, \cdots, 0)^T\in\mathbb{R}^p$, $\bm{\Sigma} = (\sigma_{ij})$ where $\sigma_{ij}=1$ for $i=j$, $\sigma_{ij}=-0.2$ for $1\le i\ne j\le 5$ and $\sigma_{ij}=0$ otherwise. Set $\bm{\beta}^0 = (1.39, 1.47, 1.56, 1.65, 1.74,0,\cdots,0)^T\in\mathbb{R}^p$ such that the Bayes rule of $\mathrm{sign}(\bm{x}^T\bm{\beta}^0)$ has Bayes error 6.3\%.

{\em Scheme 2}: $\bm{x} \sim \mathrm{N}_p(\bm{0}_p,\bm{\Sigma})$, $\bm{\Sigma} = (\sigma_{ij})$ where $\sigma_{ij}=0.4^{|i-j|}$ for $1\le i,j\le p$, $\bm{\beta}^0 = (1.1, 1.1, 1.1, 1.1, 1.1, 0,\cdots,0)^T\in\mathbb{R}^p$, $P(y=1|\bm{x}) = \Phi(\bm{x}^T\bm{\beta}^0)$ where $\Phi(\cdot)$ denotes the cumulative density function of centered t-distribution with degree of freedom 2.

{\em Scheme 3}: $\bm{x} \sim \mathrm{N}_p(\bm{0}_p,\bm{\Sigma})$, $\bm{\Sigma} = (\sigma_{ij})$ where $\sigma_{ij}=0.4^{|i-j|}$ for $1\le i,j\le p$, $\bm{\beta}^0 = (1.1, 1.1, 1.1, 1.1, 1.1, 0,\cdots,0)^T\in\mathbb{R}^p$, $P(y=1|\bm{x}) = [1+\exp(-\bm{x}^T\bm{\beta}^0)]^{-1}$.

Schemes 1 and 2 are adapted from experiments in \cite{peng2016error}. Scheme 1 resembles the linear discriminate analysis (LDA). Scheme 2 is based on a robit regression framework \citep{liu2004robit}, while Scheme 3 is in a typical logistic regression setting. In all three schemes, the predictors are designed to be correlated to certain degree. We also generate the error-prone predictors as $\bm{w} = \bm{x}+\bm{u}$ where $\bm{u} \sim \mathrm{N}_p(\bm{0}_p, \sigma_u^2\mathrm{I}_p)$. All methods use the error-prone predictors in the estimation procedure instead of the true predictors $\bm{x}$'s. We consider two levels of standard deviation $\sigma_u$ for the measurement errors at 0.3 and 0.5.

We set the training and testing sample sizes both at $n=100$ and the total number of predictors at $p=1000$ and $p=5000$. For the scenario with $p=5000$, the methods based on linear programming including GDS, GMUS and our MUC methods, cannot be applied directly for such high-dimensional problem. In this case, GDS and GMUS methods are removed from comparison. In addition to the Lasso analog approach, we also implement the proposed MUC methods for the $p=5000$ case using a preliminary screening step that first uses the Lasso analog method to retain the top 1000 predictors (in terms of absolute estimated values) and subsequently applies the MUC approach directly. We denote this approach as the hybrid procedure, with `hybrid.logistic' and `hybrid.smhinge' denoting the logistic loss and smooth hinge loss respectively.

The performance of all methods are evaluated on: (1) variable selection accuracy measured by number of false negatives (`FN') and number of false positives (`FP'); (2) estimation error in $L_1$ norm (`L1error') on the regression coefficients; (3) classification accuracy (`Accuracy') and F1 score (`F1') on the testing samples. Accuracy refers to the proportion of correctly classified labels, while the F1 score represents the harmonic mean between precision and recall (sensitivity) that capture the operating characteristics of the classifier. Both these metrics lie between 0 and 1, with a higher value indicative of a superior performance.

{\noindent \underline{\bf Results:}} Table \ref{tab:sims_class} summarizes the classification results from our proposed methods and all the competing methods. In general, the methods ignoring the measurement errors including L1logistic, $L_1$-norm SVM and GDS have considerably worse feature selection (inflated false positives) and poor coefficient estimation performance compared to the methods with noise correction strategy, particularly the proposed MUC and its variant analog methods. Such findings are supported by the fact that there is risk of admitting many false positives when ignoring measurement errors in the estimation \citep{sorensen2015measurement}. The proposed MUC approaches also consistently produce improved or similar classification accuracy and F1 scores compared to the competing methods. While the coefficient estimation performance is guaranteed by our theoretical guarantees, the good prediction performance under the proposed MUC variants is reassuring. The GMUS approach, which is the only other competing method with noise correction, consistently shows inferior coefficient estimation performance. However, it often has comparable classification performance.

\begin{table}[!htp]
\caption{Results from Classification Simulation Schemes}
\label{tab:sims_class}
\vspace{.05in}
\centering
\resizebox{.8\textwidth}{!}{
\begin{tabular}{|l|ccccc|ccccc|}
\hline
  & \textit{FN} & \textit{FP} & \textit{L1error} & \textit{Accuracy} & \textit{F1} & \textit{FN} & \textit{FP} & \textit{L1error} & \textit{Accuracy} & \textit{F1}\\
\hline
\textit{Scheme 1}  & \multicolumn{5}{c|}{$p=1000$, $\sigma_u = 0.3$} & \multicolumn{5}{c|}{$p=1000$, $\sigma_u = 0.5$} \\
\hline
L1logistic & 2 & 12 & 8.44 & 0.73 & 0.72 & 2.0 & 12.5 & 8.76 & 0.69 & 0.68 \\

L1SVM & 2 & 31 & 9.34 & 0.70 & 0.69 & 2.0 & 32.5 & 9.76 & 0.62 & 0.60 \\

GDS & 1 & 9 & 7.54 & 0.74 & 0.72 & 2.0 & 8.5 & 7.96 & 0.68 & 0.66 \\

GMUS & 2 & 12 & 7.47 & 0.73 & 0.72 & 2.0 & 9.5 & 7.73 & 0.70 & 0.69 \\

analog.logistic & 3 & 1 & 7.71 & 0.65 & 0.63 & 3.0 & 1.0 & 7.62 & 0.64 & 0.61 \\

analog.smhinge & 3 & 1 & 7.12 & 0.53 & 0.68 & 3.0 & 1.0 & 7.01 & 0.52 & 0.67 \\

muc.logistic & 2 & 1 & 7.35 & 0.73 & 0.73 & 2.5 & 2.0 & 7.58 & 0.71 & 0.70 \\

muc.smhinge & 2 & 3 & 7.14 & 0.74 & 0.74 & 2.0 & 2.0 & 7.50 & 0.70 & 0.70 \\

\hline
\textit{Scheme 1}  & \multicolumn{5}{c|}{$p=5000$, $\sigma_u = 0.3$} & \multicolumn{5}{c|}{$p=5000$, $\sigma_u = 0.5$} \\
\hline
L1logistic & 2 & 16.0 & 9.25 & 0.67 & 0.66 & 3 & 18.5 & 9.31 & 0.63 & 0.61 \\

L1SVM & 2 & 40.0 & 10.12 & 0.63 & 0.62 & 3 & 43.5 & 9.91 & 0.61 & 0.58 \\

analog.logistic & 3 & 4.5 & 8.13 & 0.60 & 0.58 & 4 & 7.0 & 8.26 & 0.58 & 0.53 \\

analog.smhinge & 3 & 4.0 & 8.51 & 0.55 & 0.68 & 4 & 6.5 & 8.84 & 0.53 & 0.65 \\

hybrid.logistic & 3 & 4.0 & 7.92 & 0.69 & 0.69 & 3 & 2.0 & 8.07 & 0.64 & 0.66 \\

hybrid.smhinge & 3 & 6.0 & 8.18 & 0.68 & 0.68 & 3 & 7.0 & 8.36 & 0.64 & 0.64\\
\hline
\textit{Scheme 2}  & \multicolumn{5}{c|}{$p=1000$, $\sigma_u = 0.3$} & \multicolumn{5}{c|}{$p=1000$, $\sigma_u = 0.5$} \\
\hline
L1logistic & 1.0 & 11 & 5.50 & 0.76 & 0.76 & 1 & 8.0 & 5.72 & 0.73 & 0.72 \\

L1SVM & 2.0 & 30 & 6.99 & 0.69 & 0.69 & 2 & 33.0 & 7.37 & 0.66 & 0.65 \\

GDS & 1.0 & 4 & 4.84 & 0.75 & 0.75 & 1 & 4.0 & 5.21 & 0.72 & 0.71 \\

GMUS & 1.0 & 5 & 4.76 & 0.76 & 0.76 & 1 & 7.0 & 5.16 & 0.73 & 0.72 \\

analog.logistic & 2.5 & 0 & 4.30 & 0.72 & 0.71 & 2 & 0.0 & 4.53 & 0.72 & 0.71 \\

analog.smhinge & 2.0 & 0 & 3.27 & 0.59 & 0.71 & 2 & 0.0 & 3.28 & 0.58 & 0.70 \\

muc.logistic & 1.0 & 3 & 4.77 & 0.77 & 0.77 & 2 & 2.0 & 5.00 & 0.73 & 0.73 \\

muc.smhinge & 1.0 & 3 & 4.33 & 0.77 & 0.77 & 1 & 2.5 & 4.80 & 0.74 & 0.73 \\
\hline
\textit{Scheme 2}  & \multicolumn{5}{c|}{$p=5000$, $\sigma_u = 0.3$} & \multicolumn{5}{c|}{$p=5000$, $\sigma_u = 0.5$} \\
\hline
L1logistic & 1 & 13.5 & 5.79 & 0.76 & 0.76 & 1.0 & 11.5 & 5.88 & 0.71 & 0.70\\

L1SVM & 2 & 39.0 & 7.23 & 0.69 & 0.69 & 2.0 & 37.0 & 7.21 & 0.63 & 0.62\\

analog.logistic & 2 & 0.0 & 5.24 & 0.74 & 0.72 & 2.0 & 0.0 & 5.35 & 0.71 & 0.69 \\

analog.smhinge & 2 & 0.0 & 4.81 & 0.63 & 0.73 & 2.0 & 0.0 & 5.04 & 0.60 & 0.71 \\

hybrid.logistic & 1 & 5.0 & 5.12 & 0.76 & 0.76 & 2.0 & 1.0 & 5.17 & 0.72 & 0.72\\

hybrid.smhinge & 1 & 5.0 & 4.86 & 0.77 & 0.77 & 2.0 & 2.0 & 5.11 & 0.72 & 0.72\\
\hline
\textit{Scheme 3}  & \multicolumn{5}{c|}{$p=1000$, $\sigma_u = 0.3$} & \multicolumn{5}{c|}{$p=1000$, $\sigma_u = 0.5$} \\
\hline
L1logistic & 1 & 3 & 5.35 & 0.76 & 0.75 & 1.0 & 4 & 5.76 & 0.72 & 0.72 \\

L1SVM & 1 & 32 & 7.27 & 0.69 & 0.69 & 1.0 & 31 & 7.18 & 0.67 & 0.67 \\

GDS & 1 & 8 & 5.02 & 0.75 & 0.76 & 1.0 & 4 & 5.17 & 0.73 & 0.72 \\

GMUS & 1 & 7 & 4.85 & 0.76 & 0.76 & 1.0 & 6 & 5.13 & 0.74 & 0.73 \\

analog.logistic & 2 & 0 & 4.30 & 0.74 & 0.75 & 2.5 & 0 & 4.61 & 0.72 & 0.71 \\

analog.smhinge & 2 & 0 & 3.44 & 0.58 & 0.70 & 2.5 & 0 & 3.51 & 0.58 & 0.70 \\

muc.logistic & 1 & 0 & 4.85 & 0.77 & 0.77 & 1.0 & 1 & 5.01 & 0.74 & 0.74 \\

muc.smhinge & 1 & 1 & 4.49 & 0.78 & 0.77 & 1.0 & 2 & 4.78 & 0.74 & 0.74 \\
\hline
\textit{Scheme 3}  & \multicolumn{5}{c|}{$p=5000$, $\sigma_u = 0.3$} & \multicolumn{5}{c|}{$p=5000$, $\sigma_u = 0.5$} \\
\hline
L1logistic & 1 & 9 & 5.65 & 0.73 & 0.72 & 1 & 6 & 5.74 & 0.73 & 0.71 \\

L1SVM & 2 & 40 & 8.40 & 0.66 & 0.66 & 2 & 38 & 8.01 & 0.66 & 0.65 \\

analog.logistic & 2 & 0 & 5.27 & 0.71 & 0.70 & 2 & 1 & 5.39 & 0.69 & 0.67 \\

analog.smhinge & 2 & 0 & 4.89 & 0.62 & 0.72 & 2 & 1 & 5.16 & 0.61 & 0.71 \\

hybrid.logistic & 1 & 2 & 5.16 & 0.74 & 0.74 & 2 & 5 & 5.26 & 0.74 & 0.74 \\

hybrid.smhinge & 1 & 3 & 5.06 & 0.74 & 0.73 & 2 & 5 & 5.19 & 0.74 & 0.74 \\
\hline
\end{tabular}}
\end{table}

Among the methods involving in-built noise correction strategy, the muc.smhinge method has generally better coefficient estimation performance compared to GMUS method as well as the muc.logistic approach, except for a few cases in Scheme 1. Remarkably, the muc.smhinge method shows improvements in coefficient estimation over muc.logistic in Scheme 3 where the data generation follows logistic regression. In terms of prediction, both the MUC variants have similar performance. The feature selection and coefficient estimation performance of the Lasso analog method with smooth hinge loss is generally superior compared to all approaches in most cases, barring a few settings in Scheme 1. However, the classification performance of the analog approaches may not be optimal in the settings that we consider.

\vspace{-.15in}
\subsection{Quantile Regression Problems}
\label{subsec:sims_quant}
We compare the empirical performance of our proposed methods with smooth quantile regression loss with the recently proposed `conquer' method from \cite{he2021smoothed}. We denote the proposed method for quantile regression as mu.qr and the analog version as analog.qr. For the $p=5000$ case, we fit the hybrid version of the matrix uncertainty estimator (similar to the approach mentioned in Section \ref{subsec:sims_class}), and denote this approach as hybrid.qr. The performance of the methods is evaluated in terms of variable selection accuracy (`FN', `FP') and coefficient estimation error (`L1error'), as well as prediction accuracy on testing samples measured by the check loss (`Check') \citep{wang2012quantile}.

Data is generated from a linear heterogeneous model scheme: $\bm{x} \sim \mathrm{N}_p(\bm{0}_p,\bm{\Sigma})$, $\bm{\Sigma} = (\sigma_{ij})$ where $\sigma_{ij}=0.7^{|i-j|}$ for $1\le i,j\le p$, $\bm{\beta}^0 = (1.5, 1.5, 1.5, 1.5, 1.5, 0,\cdots,0)^T\in\mathbb{R}^p$, $y = 1.5 + \bm{x}^T\bm{\beta}^0 + (x_1 + 1)\{\epsilon - F^{-1}_{\epsilon}(\tau)\}$. Here $F_\epsilon^{-1}(\tau)$ denotes the inverse cumulative density function of the random noise $\epsilon$ at quantile $\tau$. We generate the random noise $\epsilon$: (1) normal distribution N(0,4); (2) centered t-distribution with degree of freedom 2. This data generation scheme is similar to the one in \cite{he2021smoothed}. In addition, we generate the noisy predictors as $\bm{w} = \bm{x}+\bm{u}$ where $\bm{u} \sim \mathrm{N}_p(\bm{0}_p, \sigma_u^2\mathrm{I}_p)$, which would be used for model fitting instead of the true predictors. Two different noise levels are investigated ($\sigma_u=0.3,0.5$), with training and testing sample sizes both setting at $n=100$.

\begin{table}[!t]
\caption{Results from Quantile Regression Simulations}
\label{tab:sims_qr}
\vspace{.05in}
\centering
\resizebox{.8\textwidth}{!}{
\begin{tabular}{|l|cccc|cccc|cccc|}
\hline
  & \textit{FN} & \textit{FP} & \textit{L1error} & \textit{Check} & \textit{FN} & \textit{FP} & \textit{L1error} & \textit{Check} & \textit{FN} & \textit{FP} & \textit{L1error} & \textit{Check} \\
\hline
p=1000  & \multicolumn{4}{c|}{normal, $\tau = 0.1$} & \multicolumn{4}{c|}{normal, $\tau = 0.5$} & \multicolumn{4}{c|}{normal, $\tau = 0.9$} \\
\hline
  & \multicolumn{4}{c|}{{\em $\sigma_u=0.3$}} & \multicolumn{4}{c|}{{\em $\sigma_u=0.3$}} & \multicolumn{4}{c|}{{\em $\sigma_u=0.3$}} \\

conquer & 0 & 6 & 3.91 & 0.70 & 0 & 9.5 & 3.37 & 1.20 & 1.5 & 5.5 & 5.80 & 0.73 \\

mu.qr & 0 & 0 & 3.22 & 0.65 & 0 & 0 & 3.09 & 1.25 & 2 & 1 & 5.70 & 0.68 \\

analog.qr & 0 & 0 & 4.15 & 1.08 & 0 & 0.5 & 4.51 & 1.66 & 1 & 1 & 5.89 & 0.89 \\

  & \multicolumn{4}{c|}{{\em $\sigma_u=0.5$}} & \multicolumn{4}{c|}{{\em $\sigma_u=0.5$}} & \multicolumn{4}{c|}{{\em $\sigma_u=0.5$}} \\

conquer & 0 & 6 & 4.53 & 0.83 & 0 & 14.5 & 4.39 & 1.38 & 1 & 6 & 5.56 & 0.69 \\

mu.qr & 0.5 & 0 & 3.34 & 0.74 & 0 & 1 & 3.31 & 1.40 & 1 & 0.5 & 4.22 & 0.63 \\

analog.qr & 0 & 0 & 4.74 & 1.33 & 0 & 0 & 5.04 & 1.80 & 1 & 2.5 & 6.27 & 0.85 \\

\hline
p=1000  & \multicolumn{4}{c|}{student-t, $\tau=0.1$} & \multicolumn{4}{c|}{student-t, $\tau=0.5$} & \multicolumn{4}{c|}{student-t, $\tau=0.9$} \\
\hline

  & \multicolumn{4}{c|}{{\em $\sigma_u=0.3$}} & \multicolumn{4}{c|}{{\em $\sigma_u=0.3$}} & \multicolumn{4}{c|}{{\em $\sigma_u=0.3$}} \\

conquer & 0 & 8.5 & 5.11 & 0.72 & 0 & 8 & 2.67 & 1.11 & 1 & 6.5 & 5.51 & 0.72 \\

mu.qr & 1 & 0 & 3.58 & 0.64 & 0 & 0 & 5.74 & 1.65 & 1 & 0 & 4.43 & 0.69 \\

analog.qr & 0 & 0 & 4.51 & 1.03 & 0 & 1 & 4.76 & 1.75 & 1 & 0.5 & 5.23 & 0.83 \\

  & \multicolumn{4}{c|}{{\em $\sigma_u=0.5$}} & \multicolumn{4}{c|}{{\em $\sigma_u=0.5$}} & \multicolumn{4}{c|}{{\em $\sigma_u=0.5$}} \\

conquer & 0 & 7.5 & 4.64 & 0.82 & 0 & 13 & 3.92 & 1.32 & 1 & 4 & 5.47 & 0.73 \\

mu.qr & 0 & 0 & 3.17 & 0.75 & 0 & 0 & 3.86 & 1.40 & 1 & 0 & 4.04 & 0.65 \\

analog.qr & 0 & 0 & 4.97 & 1.23 & 0 & 1 & 5.15 & 1.87 & 1 & 2 & 6.10 & 0.84 \\

\hline
p=5000  & \multicolumn{4}{c|}{normal, $\tau = 0.1$} & \multicolumn{4}{c|}{normal, $\tau = 0.5$} & \multicolumn{4}{c|}{normal, $\tau = 0.9$} \\
\hline
  & \multicolumn{4}{c|}{{\em $\sigma_u=0.3$}} & \multicolumn{4}{c|}{{\em $\sigma_u=0.3$}} & \multicolumn{4}{c|}{{\em $\sigma_u=0.3$}} \\

conquer & 0 & 8.5 & 4.54 & 0.70 & 0 & 10.5 & 3.66 & 1.29 & 3 & 5.0 & 6.58 & 0.72 \\

hybrid.qr & 1 & 0 & 3.38 & 0.63 & 0 & 0 & 5.45 & 1.47 & 2 & 0 & 5.05 & 0.62 \\

analog.qr & 0 & 0 & 3.94 & 1.05 & 0 & 1 & 4.68 & 1.68 & 1 & 1 & 5.86 & 0.89 \\

  & \multicolumn{4}{c|}{{\em $\sigma_u=0.5$}} & \multicolumn{4}{c|}{{\em $\sigma_u=0.5$}} & \multicolumn{4}{c|}{{\em $\sigma_u=0.5$}}\\

conquer & 0 & 10.0 & 4.51 & 0.81 & 0 & 17.5 & 5.00 & 1.49 & 2 & 7.0 & 6.63 & 0.76 \\

hybrid.qr & 0.5 & 0 & 3.36 & 0.72 & 0 & 3 & 6.78 & 1.65 & 2 & 1 & 6.06 & 0.69 \\

analog.qr & 0 & 0 & 4.53 & 1.28 & 0 & 0.5 & 4.95 & 1.78 & 1 & 3.5 & 6.47 & 0.86 \\

\hline
p=5000  & \multicolumn{4}{c|}{student-t, $\tau=0.1$} & \multicolumn{4}{c|}{student-t, $\tau=0.5$} & \multicolumn{4}{c|}{student-t, $\tau=0.9$} \\
\hline

  & \multicolumn{4}{c|}{{\em $\sigma_u=0.3$}} & \multicolumn{4}{c|}{{\em $\sigma_u=0.3$}} & \multicolumn{4}{c|}{{\em $\sigma_u=0.3$}} \\

conquer & 0.0 & 9.0 & 5.21 & 0.79 & 0.0 & 21.0 & 3.44 & 1.19 & 1.5 & 8.5 & 5.88 & 0.84 \\

hybrid.qr & 0 & 0 & 3.16 & 0.67 & 0 & 0 & 2.5 & 1.21 & 1 & 0 & 3.93 & 0.73 \\

analog.qr & 0 & 0.5 & 4.34 & 0.97 & 0 & 1 & 4.93 & 1.75 & 1 & 1 & 5.39 & 0.82 \\

  & \multicolumn{4}{c|}{{\em $\sigma_u=0.5$}} & \multicolumn{4}{c|}{{\em $\sigma_u=0.5$}} & \multicolumn{4}{c|}{{\em $\sigma_u=0.5$}} \\

conquer & 0 & 8.0 & 5.14 & 0.90 & 0 & 20.0 & 4.28 & 1.31 & 2 & 9.5 & 6.18 & 0.91 \\

hybrid.qr & 0 & 0 & 3.46 & 0.78 & 0 & 1.5 & 5.42 & 1.51 & 2 & 1 & 4.66 & 0.79 \\

analog.qr & 0 & 0.5 & 4.86 & 1.21 & 0 & 2 & 5.75 & 1.89 & 1 & 1 & 5.95 & 0.86 \\

\hline
\end{tabular}}
\end{table}

{\noindent \underline{\bf Results:}} The results for quantiles at 0.1, 0.5 and 0.9 are shown in Table \ref{tab:sims_qr}. As seen from the results, the MU estimator consistently results in negligible FP and FN rates, and records considerable improvement in terms of coefficient estimation and prediction across all noise levels and quantiles considered, for $p=1000$ case. For the case with $p=5000$, the propose estimators still consistently show negligible FP and FN rates compared to the conquer approach that registers inflated FP rates. The coefficient estimation and prediction performance under the proposed method is typically superior for the tail quantiles, but may sometimes be inferior to the conquer method near the median of the distribution. Overall, the proposed estimators provides consistent improvements in feature selection by controlling the FP and FN rates compared to the conquer approach. While the coefficient estimation and prediction performance somewhat deteriorates for the $p=5000$ case, the proposed approach still registers considerable benefits towards the tails of the distribution.

\vspace{-.25in}
\section{HCP Gender Classification Application}
\label{sec:application}
We apply the proposed approach to a gender classification task in HCP data.

{\noindent \underline{\bf Data Description:}} We used the resting-state fMRI data from the FIX pre-processed left-right phase-encoding scan from the first session. The fMRI time series were processed into 360 regions based on the Glasser atlas. For details of preprocessing and information on the Glasser atlas, please see \cite{smith2013resting} and \cite{glasser2016multi}. The total sample size is 1003, with 534 female and 469 male participants.

We calculated sparse inverse covariance matrices for the 360 regions for each individual using the graphical Lasso algorithm \citep{friedman2008sparse}, which comprised a total of $360\times(360-1)/2=64620$ edges. The shrinkage parameter was tuned such that the computed inverse covariance matrix represents a network of density of approximately 0.2 across all samples. We further conducted a screening on the edges and removed those with a standard deviation less than 0.01 from our predictor list, which were non-informative for our analysis. This resulted in a final edge set of length 13073 for the analysis. Partial correlations (representing the edge strengths) corresponding to these 13073 edges were used as features in the classification model. We evaluate classification performance as well as the test-retest reliabilty of the selected features, for the proposed approach and competing methods. High test-retest reliability indicates consistent feature selection, the importance of which has been emphasized in recent work \citep{tian2021machine} for neuroimaging studies.

{\noindent \underline{\bf Results Summary}}
\label{subsec:hcp_gender}
Table \ref{tab:hcp_gender} summarizes the results from 20 random splits of data samples into training and testing sets (train/test ratio at 4:1). The average classification accuracy and F1 score on the testing set are reported. The L1logistic, L1SVM, and our analog methods with logistic and smooth hinge losses use the whole edge set of length 13073 as input, while the methods based on linear programming, including GDS, GMUS and the hybrid implementation of our MUC methods (hybrid.logistic and hybrid.smhinge), use only a retained edge set of length 500 from the analog estimation. In general, the methods using the whole edge set have superior prediction performance compared to those only using the retained shorter edge set. Overall, the proposed analog methods achieve the best prediction performance across all methods, which is significantly better than all competing approaches.

For evaluating the test-retest reliability for the selected features, we calculated pairwise Pearson's correlations between the estimated coefficient vectors across all the 20 replicates for each method. A robust approach is expected to demonstrate high test-retest reliability by consistently selecting the same set of  important features across all replicates, which should translate to a high correlation for the estimated regression coefficients across replicates. The summary of correlations is reported in Table \ref{tab:hcp_gender}, and illustrates considerably higher test-retest reliability ($\ge 0.8$) under the proposed analog approach compared to all competing approaches. L1logistic and L1SVM have worse reliability compared to analog methods, with correlations below 0.65 in the best case settings. The other methods using only the retained top edges share the worst reliability measures. Importantly, the minimum correlations across all replicates under the competing approaches are orders of magnitudes lower compared to the proposed method, indicating weak agreement of the selected features between some replicates. These results together with the prediction performance, these results demonstrate the advantage of including an exhaustive set of edges for analysis, and the importance of accounting for noise for robust feature selection in high-dimensional analysis.

We also summarize the number of selected edges for each method in Table \ref{tab:hcp_gender}. For each edge, we perform a one sample t-test for the 20 estimates obtained from the 20 randomly selected training sets. We report the number of significant edges after correction for multiple testing via Bonferroni method. The proposed analog methods using whole edge set are among the methods that select the least number of significant edges, showing the ability to effectively control the selection of false positives. On the other hand, the L1logistic and L1SVM methods using the whole edge set select the most number of edges, pointing to the possibility of including many false positive signals based on our experience with extensive simulation studies in Section \ref{sec:sims}. Out of the 23 and 25 edges selected by the analog approaches using the logistic and smooth hinge loss respectively, 17 common edges were selected by both approaches. Out of these 17 edges, 15 of them were included in the selection of L1logistic method, and 14 were included for L1SVM method. This indicates that the significant edges identified under the proposed methods are also consistently selected by competing approaches; however these latter methods have the tendency to report inflated number of significant edges, potentially involving false positives.

\begin{table}[!t]
\caption{Analysis Summary on HCP gender classification}
\label{tab:hcp_gender}
\centering
\resizebox{.8\textwidth}{!}{
\begin{tabular}{|l|c|c|c|ccc|}
\hline
  & Accuracy & F1 & \# of edges & Corr.min & Corr.mean & Corr.max \\
\hline
L1logistic & 0.82 & 0.83 & 44 & 0.34 & 0.50 & 0.64\\

L1SVM & 0.86 & 0.87 & 113 & 0.48 & 0.55 & 0.60\\

GDS* & 0.75 & 0.79 & 23 & 0.18 & 0.30 & 0.45\\

GMUS* & 0.70 & 0.77 & 36 & 0.26 & 0.43 & 0.62\\

hybrid.logistic* & 0.77 & 0.80 & 38 & 0.23 & 0.32 & 0.43\\

hybrid.smhinge* & 0.76 & 0.79 & 23 & 0.17 & 0.29 & 0.44\\

analog.logistic & 0.90 & 0.91 & 23 & 0.80 & 0.82 & 0.84\\

analog.smhinge & 0.89 & 0.90 & 25 & 0.80 & 0.82 & 0.84\\

\hline
\end{tabular}
}

*: methods using retained top edges from analog estimation.
\end{table}

\begin{figure}[!t]
\caption{Circular plot for significant edges with gender differences}
\label{fig:circ_gender}
\hspace{.4in}\includegraphics[width=0.8\textwidth]{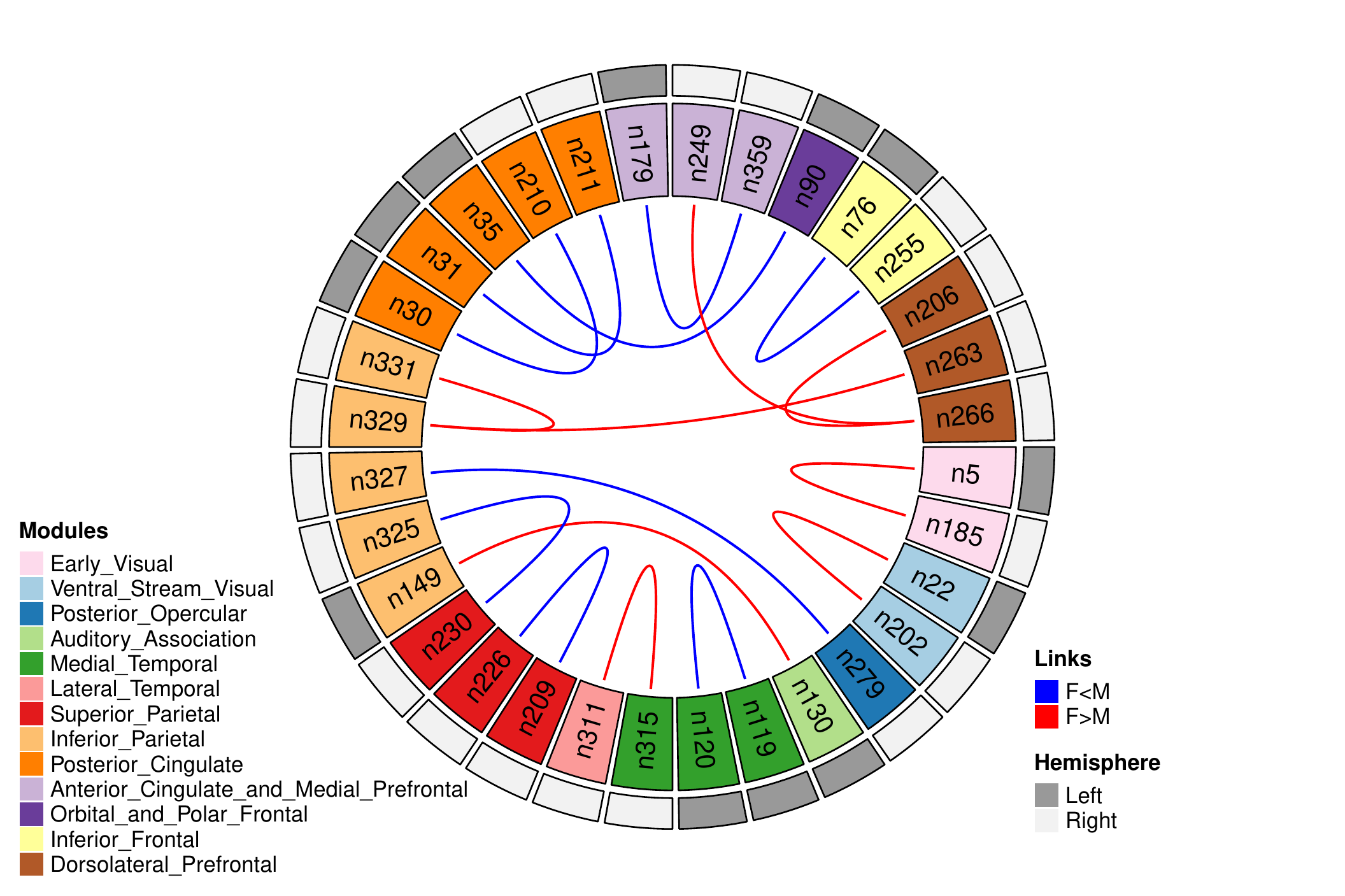}
\end{figure}

We visually illustrate the 17 common significant edges that show gender differences under both the analog methods in the circular plot in Fig \ref{fig:circ_gender}. The plot includes the region index in the Glasser atlas, and module and hemisphere information of the related cortical surface regions. We see that a subset of edges link the symmetrical brain regions between the two hemispheres, including the edges between node 5 and 185, 22 and 202, 30 and 210, 31 and 211, 179 and 359. Some remaining edges connect different functional modules, typically corresponding to regions from the same hemisphere, such as the edges between node 35 and 90, 130 and 149, 263 and 329, 279 and 327. The brain regions with the greatest number of differential edges include the posterior cingulate with edges that were more prominent in males, the dorsolateral prefrontal region with edges that were more prominent in females, and inferior parietal region containing both types of edges. These regions have been previously shown to be associated with gender differences \citep{sie2019gender,weis2020sex}.

\vspace{-.25in}
\section{Discussions}
\label{sec:discuss}
In this paper, we proposed a unified regression framework for the class of smooth Lipschitz loss functions under high-dimensional settings that is able to account for measurement errors in the predictors. The proposed estimators are able to achieve accurate coefficient estimation and feature selection without the need of knowledge on the noise distributions that expedites computation and implementation. We also propose a Lasso-type analog estimator that is scalable to higher dimensions and computationally more efficient. The proposed approach is expected to have important implications for a wide range of loss functions, including the routinely encountered classification and quantile regression problems, in the presence of measurement errors for high-dimensional covariates. The proposed methods are particularly relevant for medical imaging applications where the observed images and derived imaging features are known to be inherently contaminated with noise. This is evident given the clear advantages in classification and feature selection demonstrated by the proposed method in the brain network based gender classification task involving HCP data analysis.  The HCP analysis results demonstrating the clear advantages after corrections for measurement error are particularly relevant, given that existing prediction and classification approaches based on brain connectivity features have essentially ignored the presence of noise.  Future work will focus on extending the framework to accommodate functional covariates, and tensor-structured covariates. Potential generalizations to multi-task learning will also be considered.

\spacingset{1}
\bibliographystyle{apalike}
\bibliography{literature}

\begin{thebibliography}{}

\bibitem[Becker et~al., 2009]{becker2009penalizedsvm}
Becker, N., Werft, W., Toedt, G., Lichter, P., and Benner, A. (2009).
\newblock penalizedsvm: a r-package for feature selection svm classification.
\newblock {\em Bioinformatics}, 25(13):1711--1712.

\bibitem[Bickel et~al., 2009]{bickel2009simultaneous}
Bickel, P.~J., Ritov, Y., and Tsybakov, A.~B. (2009).
\newblock Simultaneous analysis of lasso and dantzig selector.
\newblock {\em Ann. Stat.}, 37(4):1705--1732.

\bibitem[Birgin et~al., 2000]{birgin2000nonmonotone}
Birgin, E.~G., Mart{\'\i}nez, J.~M., and Raydan, M. (2000).
\newblock Nonmonotone spectral projected gradient methods on convex sets.
\newblock {\em SIAM J. Optim}, 10(4):1196--1211.

\bibitem[Candes and Tao, 2007]{candes2007dantzig}
Candes, E. and Tao, T. (2007).
\newblock The dantzig selector: Statistical estimation when p is much larger
  than n.
\newblock {\em Ann. Stat.}, 35(6):2313--2351.

\bibitem[Carroll and Stefanski, 1994]{carroll1994measurement}
Carroll, R.~J. and Stefanski, L.~A. (1994).
\newblock Measurement error, instrumental variables and corrections for
  attenuation with applications to meta-analyses.
\newblock {\em Stat Med.}, 13(12):1265--1282.

\bibitem[Chen and Caramanis, 2013]{chen2013noisy}
Chen, Y. and Caramanis, C. (2013).
\newblock Noisy and missing data regression: Distribution-oblivious support
  recovery.
\newblock In {\em International Conference on Machine Learning}, pages
  383--391. PMLR.

\bibitem[Datta and Zou, 2017]{datta2017cocolasso}
Datta, A. and Zou, H. (2017).
\newblock Cocolasso for high-dimensional error-in-variables regression.
\newblock {\em Ann. Stat.}, 45(6):2400--2426.

\bibitem[Dedieu, 2019]{dedieu2019sparse}
Dedieu, A. (2019).
\newblock Sparse (group) learning with lipschitz loss functions: a unified
  analysis.
\newblock {\em arXiv preprint arXiv:1910.08880}.

\bibitem[Duchi et~al., 2008]{duchi2008efficient}
Duchi, J., Shalev-Shwartz, S., et~al. (2008).
\newblock Efficient projections onto the l 1-ball for learning in high
  dimensions.
\newblock In {\em Proc. 25th Int. Conf. on ML}, pages 272--279.

\bibitem[Edelman, 1988]{edelman1988eigenvalues}
Edelman, A. (1988).
\newblock Eigenvalues and condition numbers of random matrices.
\newblock {\em SIAM journal on matrix analysis and applications},
  9(4):543--560.

\bibitem[Fan and Li, 2001]{fan2001variable}
Fan, J. and Li, R. (2001).
\newblock Variable selection via nonconcave penalized likelihood and its oracle
  properties.
\newblock {\em J Am Stat Assoc.}, 96(456):1348--1360.

\bibitem[Fan and Lv, 2011]{fan2011nonconcave}
Fan, J. and Lv, J. (2011).
\newblock Nonconcave penalized likelihood with np-dimensionality.
\newblock {\em IEEE Transactions on Information Theory}, 57(8):5467--5484.

\bibitem[Friedman et~al., 2008]{friedman2008sparse}
Friedman, J., Hastie, T., and Tibshirani, R. (2008).
\newblock Sparse inverse covariance estimation with the graphical lasso.
\newblock {\em Biostatistics}, 9(3):432--441.

\bibitem[Friedman et~al., 2010]{friedman2010regularization}
Friedman, J., Hastie, T., and Tibshirani, R. (2010).
\newblock Regularization paths for generalized linear models via coordinate
  descent.
\newblock {\em J Stat Softw.}, 33(1):1--22.

\bibitem[Glasser et~al., 2016]{glasser2016multi}
Glasser, M.~F., Coalson, T.~S., Robinson, E.~C., Hacker, C.~D., Harwell, J.,
  Yacoub, E., Ugurbil, K., Andersson, J., Beckmann, C.~F., Jenkinson, M.,
  et~al. (2016).
\newblock A multi-modal parcellation of human cerebral cortex.
\newblock {\em Nature}, 536(7615):171--178.

\bibitem[Guha and Rodriguez, 2021]{guha2021bayesian}
Guha, S. and Rodriguez, A. (2021).
\newblock Bayesian regression with undirected network predictors with an
  application to brain connectome data.
\newblock {\em J Am Stat Assoc.}, 116(534):581--593.

\bibitem[He et~al., 2021]{he2021smoothed}
He, X., Pan, X., Tan, K.~M., and Zhou, W.-X. (2021).
\newblock Smoothed quantile regression with large-scale inference.
\newblock {\em J. Econom.}

\bibitem[James and Radchenko, 2009]{james2009generalized}
James, G.~M. and Radchenko, P. (2009).
\newblock A generalized dantzig selector with shrinkage tuning.
\newblock {\em Biometrika}, 96(2):323--337.

\bibitem[Liu, 2004]{liu2004robit}
Liu, C. (2004).
\newblock Robit regression: a simple robust alternative to logistic and probit
  regression.
\newblock {\em Applied Bayesian Modeling and Casual Inference from
  Incomplete-Data Perspectives}, pages 227--238.

\bibitem[Liu, 2016]{liu2016noise}
Liu, T.~T. (2016).
\newblock Noise contributions to the fmri signal: An overview.
\newblock {\em NeuroImage}, 143:141--151.

\bibitem[Loh et~al., 2012]{loh2012high}
Loh, P.-L., Wainwright, M.~J., et~al. (2012).
\newblock High-dimensional regression with noisy and missing data: Provable
  guarantees with nonconvexity.
\newblock {\em Ann. Stat.}, 40(3):1637--1664.

\bibitem[Ma and Kundu, 2022]{ma2022multi}
Ma, X. and Kundu, S. (2022).
\newblock Multi-task learning with high-dimensional noisy images.
\newblock {\em In press, J Am Stat Assoc., Theory \&
  Methods,doi=10.1080/01621459.2022.2140052}.

\bibitem[Ma et~al., 2022]{ma2022semi}
Ma, X., Kundu, S., and Stevens, J. (2022).
\newblock Semi-parametric bayes regression with network-valued covariates.
\newblock {\em Machine Learning}, pages 1--35.

\bibitem[Negahban et~al., 2012]{negahban2012unified}
Negahban, S.~N., Ravikumar, P., Wainwright, M.~J., and Yu, B. (2012).
\newblock A unified framework for high-dimensional analysis of $ m $-estimators
  with decomposable regularizers.
\newblock {\em Statistical science}, 27(4):538--557.

\bibitem[Peng et~al., 2016]{peng2016error}
Peng, B., Wang, L., and Wu, Y. (2016).
\newblock An error bound for l1-norm support vector machine coefficients in
  ultra-high dimension.
\newblock {\em J Mach Learn Res.}, 17(1):8279--8304.

\bibitem[Raser and O'shea, 2005]{raser2005noise}
Raser, J.~M. and O'shea, E.~K. (2005).
\newblock Noise in gene expression: origins, consequences, and control.
\newblock {\em Science}, 309(5743):2010--2013.

\bibitem[Raskutti et~al., 2010]{raskutti2010restricted}
Raskutti, G., Wainwright, M.~J., and Yu, B. (2010).
\newblock Restricted eigenvalue properties for correlated gaussian designs.
\newblock {\em J Mach Learn Res.}, 11:2241--2259.

\bibitem[Reli{\'o}n et~al., 2019]{relion2019network}
Reli{\'o}n, J. D.~A., Kessler, D., Levina, E., and Taylor, S.~F. (2019).
\newblock Network classification with applications to brain connectomics.
\newblock {\em Ann. Appl. Stat.}, 13(3):1648.

\bibitem[Rosenbaum and Tsybakov, 2010]{rosenbaum2010sparse}
Rosenbaum, M. and Tsybakov, A.~B. (2010).
\newblock Sparse recovery under matrix uncertainty.
\newblock {\em Ann. Stat.}, 38(5):2620--2651.

\bibitem[Rosenbaum and Tsybakov, 2013]{rosenbaum2013improved}
Rosenbaum, M. and Tsybakov, A.~B. (2013).
\newblock Improved matrix uncertainty selector.
\newblock In {\em From Probability to Statistics and Back: High-Dimensional
  Models and Processes--A Festschrift in Honor of Jon A. Wellner}, pages
  276--290. Institute of Mathematical Statistics.

\bibitem[Sie et~al., 2019]{sie2019gender}
Sie, J.-H., Chen, Y.-H., Shiau, Y.-H., and Chu, W.-C. (2019).
\newblock Gender-and age-specific differences in resting-state functional
  connectivity of the central autonomic network in adulthood.
\newblock {\em Frontiers in human neuroscience}, 13:369.

\bibitem[Smith et~al., 2013]{smith2013resting}
Smith, S.~M., Beckmann, C.~F., Andersson, J., Auerbach, E.~J., Bijsterbosch,
  J., Douaud, G., Duff, E., Feinberg, D.~A., Griffanti, L., Harms, M.~P.,
  et~al. (2013).
\newblock Resting-state fmri in the human connectome project.
\newblock {\em Neuroimage}, 80:144--168.

\bibitem[S{\o}rensen et~al., 2015]{sorensen2015measurement}
S{\o}rensen, {\O}., Frigessi, A., and Thoresen, M. (2015).
\newblock Measurement error in lasso: Impact and likelihood bias correction.
\newblock {\em Statistica sinica}, pages 809--829.

\bibitem[S{\o}rensen et~al., 2018]{sorensen2018covariate}
S{\o}rensen, {\O}., Hellton, K.~H., Frigessi, A., and Thoresen, M. (2018).
\newblock Covariate selection in high-dimensional generalized linear models
  with measurement error.
\newblock {\em Journal of Computational and Graphical Statistics},
  27(4):739--749.

\bibitem[Stefanski, 2000]{stefanski2000measurement}
Stefanski, L.~A. (2000).
\newblock Measurement error models.
\newblock {\em J Am Stat Assoc.}, 95(452):1353--1358.

\bibitem[Tian and Zalesky, 2021]{tian2021machine}
Tian, Y. and Zalesky, A. (2021).
\newblock Machine learning prediction of cognition from functional
  connectivity: Are feature weights reliable?
\newblock {\em NeuroImage}, 245:118648.

\bibitem[Tibshirani, 1996]{tibshirani1996regression}
Tibshirani, R. (1996).
\newblock Regression shrinkage and selection via the lasso.
\newblock {\em Journal of the Royal Statistical Society: Series B
  (Methodological)}, 58(1):267--288.

\bibitem[Wang et~al., 2012a]{wang2012corrected}
Wang, H.~J., Stefanski, L.~A., and Zhu, Z. (2012a).
\newblock Corrected-loss estimation for quantile regression with covariate
  measurement errors.
\newblock {\em Biometrika}, 99(2):405--421.

\bibitem[Wang et~al., 2012b]{wang2012quantile}
Wang, L., Wu, Y., and Li, R. (2012b).
\newblock Quantile regression for analyzing heterogeneity in ultra-high
  dimension.
\newblock {\em J Am Stat Assoc.}, 107(497):214--222.

\bibitem[Wei and Carroll, 2009]{wei2009quantile}
Wei, Y. and Carroll, R.~J. (2009).
\newblock Quantile regression with measurement error.
\newblock {\em J Am Stat Assoc.}, 104(487):1129--1143.

\bibitem[Weis et~al., 2020]{weis2020sex}
Weis, S., Patil, K.~R., Hoffstaedter, F., Nostro, A., Yeo, B.~T., and Eickhoff,
  S.~B. (2020).
\newblock Sex classification by resting state brain connectivity.
\newblock {\em Cerebral cortex}, 30(2):824--835.

\bibitem[Zhang, 2010]{zhang2010nearly}
Zhang, C.-H. (2010).
\newblock Nearly unbiased variable selection under minimax concave penalty.
\newblock {\em Ann. Stat.}, 38(2):894--942.

\bibitem[Zhu et~al., 2011]{zhu2011sparsity}
Zhu, H., Leus, G., and Giannakis, G.~B. (2011).
\newblock Sparsity-cognizant total least-squares for perturbed compressive
  sampling.
\newblock {\em IEEE Trans. Signal Process}, 59(5):2002--2016.

\end{thebibliography}

\end{document}